%% file: arxiv.tex
\documentclass[12pt]{article}
\usepackage{amsmath}
\usepackage{graphicx}
\usepackage{enumerate}
\usepackage{natbib}
\usepackage{url} 
\usepackage{hyperref}

\newcommand{\blind}{1}

\usepackage{psfrag,epsf}
\usepackage{physics}

\usepackage{amsthm,color}
\usepackage{amssymb}
\usepackage{latexsym}
\usepackage{float}
\usepackage{amsfonts}
\usepackage{bbm}
\usepackage{longtable}
\usepackage{booktabs}
\usepackage{multirow}
\usepackage{mathtools}
\usepackage{relsize}

\usepackage{tikz}
\usepackage{pgfplots}
\pgfplotsset{compat = newest}
\usetikzlibrary{decorations.pathreplacing,calligraphy}
\usetikzlibrary{automata, positioning}

\usepackage{adjustbox}

\usepackage{pdfpages}
\usepackage{flafter}

\def\bD{\mbox{\boldmath $D$}}

\def\by{\mbox{\boldmath $y$}}

\def\bzeta{\mbox{\boldmath $\zeta$}}

\def\bb{\mbox{\boldmath $b$}}

\def\bY{\mbox{\boldmath $Y$}}

\def\bR{\mbox{\boldmath $R$}}
\def\bSigma{\mbox{\boldmath $\Sigma$}}

\def\bUpsilon{\mbox{\boldmath $\Upsilon$}}

\def\bGamma{\mbox{\boldmath $\Gamma$}}
\def\bPsi{\mbox{\boldmath $\Psi$}}
\def\bI{\mbox{\boldmath $I$}}

\def\bmu{\mbox{\boldmath $\mu$}}
\def\balpha{\mbox{\boldmath $\alpha$}}
\def\bpi{\vb*{\pi}}

\def\bX{\mbox{\boldmath $X$}}
\def\bbeta{\mbox{\boldmath $\beta$}}
\def\Vec{\text{vec}}
\def\bomega{\vb*{\omega}}
\def\bgamma{\vb*{\gamma}}
\def\bnu{\vb*{\nu}}

\def\bA{\vb{A}}

\usepackage{xcolor}
\usepackage[linesnumbered,ruled,vlined]{algorithm2e}

\SetCommentSty{mycommfont}
\SetKwInput{KwInput}{Input}      
\SetKwInput{KwOutput}{Output}    

\def\spacingset#1{\renewcommand{\baselinestretch}%
{#1}\small\normalsize} \spacingset{1}

\begin{document}

\input{main.tex}

\clearpage

\setcounter{section}{0}
\setcounter{subsection}{0}
\setcounter{subsubsection}{0}
\setcounter{equation}{0}
\setcounter{figure}{0}
\setcounter{table}{0}

\makeatletter
\renewcommand{\theHsection}{B.\thesection}
\renewcommand{\theHsubsection}{B.\thesubsection}
\renewcommand{\theHsubsubsection}{B.\thesubsubsection}
\renewcommand{\theHequation}{B.\theequation}
\renewcommand{\theHfigure}{B.\thefigure}
\renewcommand{\theHtable}{B.\thetable}
\makeatother

\spacingset{1}

\begin{center}
\textbf{\textit{\Huge Supplementary Materials}}\\ \vspace{3ex}
\end{center}

\input{supp.tex}

\end{document}

%% file: main.tex
\title{\bf Identification of physiological shock in intensive care units via Bayesian regime switching models}
\author{Emmett B. Kendall$^{1}$, Jonathan P. Williams$^2$, Curtis B. Storlie$^3$,\\
   Misty A. Radosevich$^3$, Erica D. Wittwer$^3$, and Matthew A. Warner$^3$\\ \\
    $^1$Department of Mathematical Sciences, The University of Texas at Dallas\\
    $^2$Department of Statistics, North Carolina State University\\
    $^3$Department of Quantitative Health Sciences, Mayo Clinic}
\date{}

\maketitle
\bigskip
\begin{abstract}
Detection of occult hemorrhage (i.e., internal bleeding) in patients in intensive care units (ICUs) can pose significant challenges for critical care workers. Because blood loss may not always be clinically apparent, clinicians rely on monitoring vital signs for specific trends indicative of a hemorrhage event. The inherent difficulties of diagnosing such an event can lead to late intervention by clinicians which has catastrophic consequences. Therefore, a methodology for early detection of hemorrhage has wide utility. We develop a Bayesian regime switching model (RSM) that analyzes trends in patients' vitals and labs to provide a probabilistic assessment of the underlying physiological state that a patient is in at any given time. This article is motivated by a comprehensive dataset we curated from Mayo Clinic of 33,924 real ICU patient encounters. Longitudinal response measurements are modeled as a vector autoregressive process conditional on all latent states up to the current time point, and the latent states follow a Markov process. We present a novel Bayesian sampling routine to learn the posterior probability distribution of the latent physiological states, as well as develop an approach to account for pre-ICU-admission physiological changes. A simulation and real case study illustrate the effectiveness of our approach.
\end{abstract}

\noindent%
\vfill

\newpage

\spacingset{1.9} 

\section{Introduction}\label{chap3:sec:intro}
Hemorrhage, especially occult blood loss, is a serious and potentially life threatening complication that is known to be difficult to diagnose. Notably, 20-40\% of hospital patients die due to injury-related-hemorrhage that was \textit{preventable} if the internal bleeding had been recognized earlier \citep{Holcomb_02}. Additionally, in those with trauma-related hemorrhage, 40\% of preventable deaths are related to inadequate hemorrhage recognition or control \citep{Holcomb_02,stensballe2017}. Patients that have experienced serious trauma and/or hypovolemic shock see an increased risk of more severe internal bleeding events \citep{moore2021trauma}. There are many difficulties in the diagnosis of bleeding, and part of the problem is the wide variation in what precisely defines a ``major bleed'' coupled with the fact that many other medical ailments can disguise the canonical behavior of hemorrhage \citep{maier2024contemporary}. This difficulty in detection can ultimately mean delayed diagnoses and care, which can lead to physiological shock and possibly death. 

The motivation for this work comes from a comprehensive dataset comprising of 33,924 patient encounters that our team of researchers has curated from Mayo Clinic's ICUs. Clinicians at Mayo Clinic are interested in a data-driven and model-based approach to reduce delay in the diagnosis of internal bleeding. The model developed should, (1) account for inter-individual differences in physiological response to hemorrhage, (2) be robust to missingness in vital and lab measurements, (3) incorporate medication information, and (4) account for pre-ICU-admission physiological changes. 

The major contribution of our work is an innovative and in-depth case study about the performance of using a Bayesian semi-supervised RSM to detect internal bleeding. Many challenges exist for this application, requiring novel statistical innovation. First, we develop a unique state-sampling routine to efficiently estimate the discrete posterior distribution of the latent physiological state for every time point for each patient encounter, both defining and estimating labels for patients' physiological conditions. Second, we fit a parametric model such that the results are interpretable to a clinician as opposed to more common, ``black-box'' machine learning (ML) approaches. Third, the mean structure for our Bayesian RSM is able to account for physiological changes prior to a patient's ICU admission. Lastly, we test the efficacy of our approach on a small test set of patient encounters that have been manually clinically annotated, to determine the accuracy of model predictions on real data. 

With advancements in computing technology and statistical learning methodology, the development of RSMs (or state-space models) for applications to the biological and medical sciences has grown  markedly \citep{kalbfleisch1985analysis, Satten1996, Bureau2003, Jackson2003, Scott2005, altman2007, Shirley2010, Zhao2016, langrock2018, li2019, williams2020bayesian, sidrow2022}. Many of these applications are with hidden Markov models (HMMs), perhaps the most fundamental example of an RSM. An HMM, and RSM more broadly, is used to model two simultaneous stochastic processes: an observed response process and a hidden state process. For a thorough review of HMMs, see \cite{rabiner1989tutorial}; for more recent applications of HMMs, see \cite{storlie2014modeling, kendall2024beyond, volpe2025prior}. The full potential of the RSM framework for biomedical research, however, remains to be realized, and this is particularly true for hierarchical Bayesian constructions of RSMs with applications for patient monitoring.  Statistical learning techniques for detecting adverse physiological events are inherently hard to train because data labels are often difficult to characterize or simply unavailable.  Dealing with a lack of gold-standard-labeled training data is a common challenge for RSM applications.  This challenge is addressed in \cite{Trabelsi2013} where, as they describe, an HMM can be trained on data to learn latent states in the absence of annotated data, in an unsupervised fashion, assuming the number of latent states to learn is known.  In this case, the HMM acts as a classification algorithm and takes into account time-series regime changes to characterize each latent state. In order to better control the model complexity through prior density specifications, we note the utilization of unsupervised Bayesian HMMs for clinical diagnoses more recently in \cite{wang2023bayesian} and \cite{lu2023bayesian}. 

While the approach we present can be considered a transparent learning algorithm, many black-box ML or artificial intelligence (AI) approaches are proving to be quite useful \citep[e.g.,][]{Hornbrook2017, bedoya2020machine, KWON2020e358, pannu2020deep, mclouth2021validation, itzhak2023prediction, jha2023grappel}. Our developed RSM extends what exists in the statistical/ML literature to build a tool that more closely addresses features of the data that are most relevant towards adequately defining and detecting patients at high risk for shock and internal bleeding.  For example, the mean structure of our response model is an approximation to how an anesthesiologist would characterize a shock event (see Section \ref{chap3:subsubsec:clinic}). Moreover, the Bayesian framework provides an interpretable approach to quantifying the likelihood of a bleeding event based on four response outcomes (heart rate, mean arterial pressure (MAP), hemoglobin, and lactate) by way of a discrete posterior distribution of a latent state sequence across time. At any point on a discretized grid of time, the model can provide a probabilistic notion of the chance of bleeding from which a clinician can then interpret and act accordingly. 

Lastly, the quality and quantity of the electronic health record data that we have gathered and curated for training and testing leads to a case study that offers real clinical insight and ramifications. Not only do these data contain vital sign and lab measurements for 33,924 patients, but they also contain detailed medication records (more in Section \ref{chap3:subsec:data}).  The data were retrospectively gathered and curated specifically for our study.

The remainder of the paper is structured as follows. Section \ref{chap3:sec:back} provides background on the medical importance of our proposed procedure (Section \ref{chap3:subsec:medRel}), a detailed description of the data (Section \ref{chap3:subsec:data}), and information on state-space models more broadly (Section \ref{chap3:subsec:stateSpace}). Section \ref{chap3:sec:method} provides the explicit model construction as well as the novel state-sampling algorithm (Section \ref{chap3:subsec:stateSamp}). Then, Section \ref{chap3:sec:sim} presents a thorough simulation study to evaluate model performance with respect to determining/calibrating the likelihood of a bleeding event. This is followed by real data results and a case study in Section \ref{chap3:sec:realDataAnalysis}. Lastly, Section \ref{chap3:sec:conc} discusses the clinical ramifications of this work as well as areas for future work.

\section{Background}\label{chap3:sec:back}
\subsection{Medical Importance} \label{chap3:subsec:medRel}
A key problem that ICUs face is that numerous patients suffer major health complications due to bleeding and shock events that go \textit{undetected} for too long. One study found that among patients admitted with severe trauma, one in three patients saw an intervention-time three or more hours after hospital admission, and one in six had an intervention-time six or more hours post-admission \citep{tran2020early}. The improvement of outcomes during acute bleeding and other shock states is primarily dependent on prompt diagnosis and management \citep{strehlow2010early}, requiring ``time sensitivity and patient specificity'' \citep{convertino2022advanced}. 
In the setting of suspected hemorrhage related to trauma, diagnosis by emergency medical providers includes a Focused Assessment with Sonography in Trauma (FAST) exam and other imaging such as computed tomography (CT) \citep{latif2023traumatic}.
Resuscitation and procedural intervention, including transfusion and surgery, are based on these findings as well as vital signs and lab work \citep{hooper2022hemorrhagic}.
In the world of trauma medicine, the ``golden hour'' concept, in which prioritization is given to rapid treatment with key interventions in the first hours of patient deterioration, is considered the gold standard in the management of trauma patients \citep{sampalis1993,sampalis1999,clarke2002}. Similarly, time critical management of patients with sepsis according to the Surviving Sepsis guidelines has led to improved outcomes for this population \citep{evans2021surviving}.
In patients with shock secondary to infection (i.e., sepsis), each 1-hour delay in antibiotic initiation is associated with a 10\% increase in mortality \citep{peltan2019}. In times of acute bleeding, recognition and treatment even minutes earlier may lead to substantial improvements in patient outcomes, as death from exsanguination can occur in as little as five minutes \citep{kotwal2018}. 

\subsection{Data Description} \label{chap3:subsec:data}
The focal point of this analysis is on the 33,924 real patient encounters from the ICUs of Mayo Clinic. This dataset consists of 33 distinct types of lab measurements, vital sign recordings, and other medical descriptors of the patients over their encounters in the ICU. The data are structured in a panel-observed format with measurements discretized to a 15 minute grid. Missing data for each patient varies depending on the specific response measurement; for example, lab measurements have a high degree of missingness, whereas vital sign recordings have little to none.

Additionally, a detailed medication history is provided for each individual. Medication information is critical for detecting internal bleeding. In particular, medications affect the trends in heart rate and MAP (not hemoglobin and lactate), and depending on the dose and/or frequency with which these drugs are administered, their effects on the trends of these vitals can be exacerbated. Therefore, by \textit{not} accounting for medications in the model, we risk confounding trends in physiological condition based on heart rate and MAP with the possible influence of medication administration.

Further details about the data cleaning process and medication information can be found in Supplementary Materials Section \ref{chap3:app:data}.

\subsection{Review of State-Space Models} \label{chap3:subsec:stateSpace}
When implementing state-space models, there usually exists two inferential interests: (1) model parameter estimates, and (2) the ``most-likely'' latent state sequence. For our purposes, the latter is of greater importance because learning the individual state sequences translates to learning the onset of bleeding or shock events. That said, the inference from the model parameters makes it possible to provide an interpretable probabilistic assessment of the likelihood of each latent state. This type of transparent learning algorithm is in contrast to the current status quo of black-box ML and AI approaches. 

\begin{figure}[!htb]
    \centering
    \spacingset{1}
    \begin{tikzpicture}[node distance={14mm}, main/.style = {draw, scale=.67}, on grid, auto] 
        \node[main] (A0) [draw=none] {$\hdots$}; 
        \node[main] (A1) [right of=A0]{$s_{k-1}$}; 
        \node[main] (A2) [right of=A1] {$s_{k}$}; 
        \node[main] (A3) [right of=A2] {$s_{k+1}$}; 
        \node[main] (A4) [draw=none, right of=A3] {$\hdots$}; 
        \node[main] (Aa0) [draw=none, below of=A0] {$\hdots$}; 
        \node[main] (Aa) [below of=A1] {$\vb*{y}_{k-1}$}; 
        \node[main] (Ab) [below of=A2] {$\vb*{y}_{k}$}; 
        \node[main] (Ac) [below of=A3] {$\vb*{y}_{k+1}$};
        \node[main] (Ad) [draw=none, below of=A4] {$\hdots$};
        \node[draw] [left of=A0] {\footnotesize hidden};
        \node[draw] [left of=Aa0] {\footnotesize observed};
        
        \draw[->] (A0) -- (A1); 
        \draw[->] (A1) -- (A2); 
        \draw[->] (A2) -- (A3);
        \draw[->] (A3) -- (A4); 
        
        \draw[->] (A1) -- (Aa); 
        \draw[->] (A2) -- (Ab);
        \draw[->] (A3) -- (Ac);
        \draw[decorate,decoration={brace,amplitude=10pt,raise=12pt}] (A0) -- (A4) node[midway,yshift=25pt] {\textbf{HMM}};
        
        \node[main] (B0) [draw=none, right=1cm of A4] {$\hdots$}; 
        \node[main] (B1) [right of=B0]{$s_{k-1}$}; 
        \node[main] (B2) [right of=B1] {$s_{k}$}; 
        \node[main] (B3) [right of=B2] {$s_{k+1}$}; 
        \node[main] (B4) [draw=none, right of=B3] {$\hdots$}; 
        \node[main] (Ba0) [draw=none, below of=B0] {$\hdots$}; 
        \node[main] (Ba) [below of=B1] {$\vb*{y}_{k-1}$}; 
        \node[main] (Bb) [below of=B2] {$\vb*{y}_{k}$}; 
        \node[main] (Bc) [below of=B3] {$\vb*{y}_{k+1}$};
        \node[main] (Bd) [draw=none, below of=B4] {$\hdots$};
        
        \draw[->] (B0) -- (B1); 
        \draw[->] (Ba0) -- (Ba); 
        \draw[->] (B1) -- (B2); 
        \draw[->] (Ba) -- (Bb); 
        \draw[->] (B2) -- (B3);
        \draw[->] (Bb) -- (Bc); 
        \draw[->] (B3) -- (B4);
        \draw[->] (Bc) -- (Bd); 
        
        \draw[->] (B1) -- (Ba); 
        \draw[->] (B2) -- (Bb);
        \draw[->] (B3) -- (Bc);
        \draw[decorate,decoration={brace,amplitude=10pt,raise=12pt}] (B0) -- (B4) node[midway,yshift=25pt] {\textbf{AR-HMM}};

        \node[main] (C0) [draw=none, right=1cm of B4] {$\hdots$}; 
        \node[main] (C1) [right of=C0]{$s_{k-1}$}; 
        \node[main] (C2) [right of=C1] {$s_{k}$}; 
        \node[main] (C3) [right of=C2] {$s_{k+1}$}; 
        \node[main] (C4) [draw=none, right of=C3] {$\hdots$}; 
        \node[main] (Ca0) [draw=none, below of=C0] {$\hdots$}; 
        \node[main] (Ca) [below of=C1] {$\vb*{y}_{k-1}$}; 
        \node[main] (Cb) [below of=C2] {$\vb*{y}_{k}$}; 
        \node[main] (Cc) [below of=C3] {$\vb*{y}_{k+1}$};
        \node[main] (Cd) [draw=none, below of=C4] {$\hdots$};
        
        \draw[->] (C0) -- (C1); 
        \draw[->] (Ca0) -- (Ca); 
        \draw[->] (C1) -- (C2); 
        \draw[->] (Ca) -- (Cb); 
        \draw[->] (C2) -- (C3);
        \draw[->] (Cb) -- (Cc); 
        \draw[->] (C3) -- (C4);
        \draw[->] (Cc) -- (Cd);

        \draw (C0) edge[out=270,in=135,looseness=.5,->] (Ca);
        \draw (C0) edge[out=285,in=135,looseness=.5,->] (Cb);
        \draw (C0) edge[out=300,in=135,looseness=.5,->] (Cc);
        \draw (C0) edge[out=315,in=135,looseness=.5,->] (Cd);

        \draw (C1) edge[out=285,in=125,looseness=.5,->] (Cb);
        \draw (C1) edge[out=300,in=125,looseness=.5,->] (Cc);
        \draw (C1) edge[out=315,in=125,looseness=.5,->] (Cd);

        \draw (C2) edge[out=300,in=115,looseness=.5,->] (Cc);
        \draw (C2) edge[out=315,in=115,looseness=.5,->] (Cd);

        \draw (C3) edge[out=315,in=105,looseness=.5,->] (Cd);
        
        \draw[->] (C1) -- (Ca); 
        \draw[->] (C2) -- (Cb);
        \draw[->] (C3) -- (Cc);
        
        \draw[decorate,decoration={brace,amplitude=10pt,raise=12pt}] (C0) -- (C4) node[midway,yshift=25pt] {\textbf{Our RSM}};
    \end{tikzpicture}
    \caption{\footnotesize Schematic of the model dependence structure for an HMM, an AR-HMM of order one, and the RSM used in our approach, respectively, from left to right. Let $\vb*{y}_k$ be some observed response vector at a time instance $k$ and $s_k$ be the corresponding latent state.}
    \label{chap3:fig:ar-hmm}
    \vspace{-1em}
\end{figure}
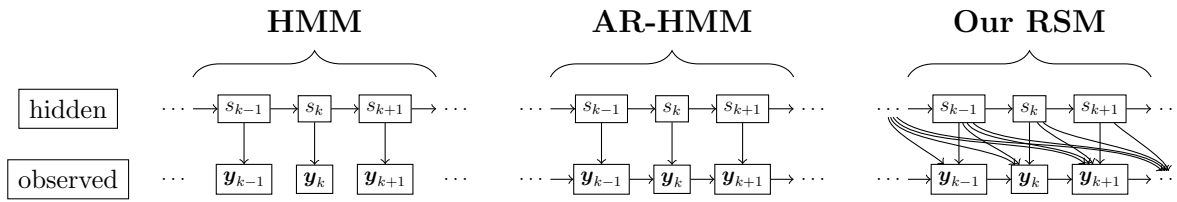

Although the HMM is the most ubiquitous form of a state-space model, its dependence structure does not adequately capture the nuanced relationship between our biological response and latent physiological states. Instead of assuming the responses are conditionally independent given the latent states (as in an HMM), we assume the responses follow an autoregressive process of order one. This describes an autoregressive HMM (AR-HMM), and applications of AR-HMMs include \cite{ailliot2012markov, stanculescu2013, williams2024bayesian}, among others. In addition to adding an autoregressive component to the response model, we also assume that the response at a given time instance is dependent on \textit{all} latent states up to that given time point (explicit justification in Section \ref{chap3:sec:method}). Therefore, rather than characterizing our model as an HMM or AR-HMM, we broadly refer to it as an RSM. There exist many examples of RSMs,  including Markov switching processes, switching autoregressive processes, and switching dynamic linear systems, among others (see \cite{puerto2021autoregressive} for an overview of the various types of RSMs).  Generally, RSMs offer more model flexibility by weakening assumptions common to HMMs. Figure \ref{chap3:fig:ar-hmm} presents a schematic of how the dependence structure differs between an HMM, an AR-HMM of order one, and our RSM.

Because of the additional dependencies between the response and hidden state process, many of the well-founded inferential strategies for HMMs and AR-HMMs breakdown or are too computationally burdensome. In particular, many inferential approaches for HMMs, or AR-HMMs alike, rely on the fact that the observed process at a given time point is \textit{only} dependent on the latent state process at that same time instant. This is not a reasonable modeling assumption in our case; hence, Section \ref{chap3:sec:method} precisely details the novel Markov chain Monte Carlo (MCMC) sampling routine we develop to simultaneously learn the posterior distributions of both the model parameters and the state sequences.

\section{Methodology}\label{chap3:sec:method}

\subsection{Clinical Influence on Model Design} \label{chap3:subsubsec:clinic}
The manner with which we construct the model to detect internal bleeding should mimic that of a clinician's own diagnosing procedure; as such, it is necessary to understand the indicators of possible hemorrhagic events. First, it is well understood that during a bleeding event, we expect hemoglobin to decrease, heart rate to increase, MAP to decrease, and lactate to increase. These trends serve as the canonical approach to diagnosing hemorrhage. Henceforth, our model needs to similarly track these trends, and we can do so by defining the mean of the response model as dependent on \textit{all} latent physiological states up to a given instance of time. Second, many patients are administered drugs to stabilize their vitals. Incorporating these medications into the model is necessary in order to distinguish trends in the vitals due to physiological changes from those due to medication administration. Lastly, training this model has the additional complication that our data lack any labels indicating patient bleeding events. However, clinical expertise suggests that for historical data, if a patient received three or more red blood cell (RBC) transfusions in a 12-hour window, then some bleeding event almost certainly occurred at some point during the patient encounter. This information can serve as a partial-labeling scheme to facilitate semi-supervised learning. All of the aforementioned clinical insights shape the model construction in Section \ref{chap3:subsec:modelConst}.  

\subsection{Model Construction}\label{chap3:subsec:modelConst}
After careful consideration with clinicians, the latent physiological state-space for our RSM comprises five states: \textit{stable} (state 1), \textit{hemorrhage} (state 2), \textit{recovery from hemorrhage} (state 3), \textit{non-bleeding event} (\textit{NBE}; state 4), and \textit{non-bleeding event recovery} (\textit{NBER}; state 5). State 1 describes the health condition of a patient with a low risk for shock or any complications from bleeding. The purpose of states 4 and 5 is to provide the RSM enough flexibility to characterize physiological conditions that are not state 1, nor are adequately described by states 2 or 3. 

\begin{figure}[!htb]
    \centering
    \spacingset{1}
    \begin{tikzpicture}[scale=1, transform shape, node distance={25mm}, thick, main/.style = {draw, circle}, baseline={-.2cm}] 
    \node[main] (1) {\textbf{1}}; 
    \node[main] (2) [below left of=1, yshift = 8mm, xshift=-6mm]{\textbf{2}}; 
    \node[main] (3) [below right of=2, yshift = 4mm, xshift=-4mm] {\textbf{3}}; 
    \node[main] (4) [below right of=1, yshift = 8mm, xshift=6mm] {\textbf{4}}; 
    \node[main] (5) [below left of=4, yshift = 4mm, xshift=4mm] {\textbf{5}}; 
    \draw[very thick, ->] (1) -- (2); 
    \draw[very thick, ->] (3) -- (1);
    \draw[very thick, ->] (1) -- (4);
    \draw[very thick, ->] (5) -- (1); 
    \draw[very thick, ->] (5) -- (2); 
    \draw[very thick, ->] (3) -- (4); 
    \draw[very thick, ->] (2) -- (3);
    \draw[very thick, ->] (3) -- (2);
    \draw[very thick, ->] (4) -- (5);
    \draw[very thick, ->] (5) -- (4); 
    \draw[very thick, ->] (4) -- (2); 
    \draw[very thick, ->] (2) -- (4); 
    \draw[very thick, ->] (1) to [out=125,in=55,looseness=5] (1); 
    \draw[very thick, ->] (2) to [out=145,in=215,looseness=5] (2); 
    \draw[very thick, ->] (3) to [out=235,in=305,looseness=5] (3);   
    \draw[very thick, ->] (4) to [out=325,in=35,looseness=5] (4); 
    \draw[very thick, ->] (5) to [out=235,in=305,looseness=5] (5);   
    \end{tikzpicture} 
    \caption{\footnotesize All allowable transitions for the five physiological states.}
    \label{chap3:fig:state_trans}
\end{figure}
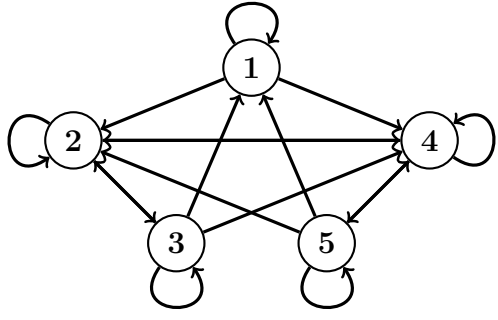
Let $N$ be the number of patients, and let $n_{i}$ be the number of time points observed for the $i^{th}$ patient, where $i \in \{1,\hdots, N\}$.  Let $\bb^{(i)}_{k}$ denote the discrete random variable corresponding to the physiological state of subject $i$ at time $k$, for $k \in \{1,\dots,n_{i}\}$, where $\bb^{(i)}_{k} \in \{1,\hdots, 5\}$. Let $\bY^{(i)}$ be a $4 \times n_{i}$ matrix with rows corresponding to the longitudinal emission variable measurements of hemoglobin, heart rate, MAP, and lactate, respectively, and columns, $\by_k^{(i)}$, corresponding to the measurements at time $k$.   

\subsubsection{Latent State Model}\label{chap3:subsubsec:latent}
Since the five physiological states are non-observable, the state sequence $\bb^{(i)}$ is most naturally treated as latent. Furthermore, because our data are observed every 15 minutes (i.e., on a grid), the state sequences are modeled as a discrete-time, discrete-state Markov process, with allowable transitions defined by Figure \ref{chap3:fig:state_trans}. The transition probability matrix, $\vb{P}$, characterizing the transitions between physiological states is defined as

{
\spacingset{1}\footnotesize
\begin{equation} \label{chap3:eq:transProb}
    \vb{P} := \mqty(\frac{1}{1 + e^{q_{1}} + e^{q_{2}}} & \frac{e^{q_{1}}}{1 + e^{q_{1}} + e^{q_{2}}} & 0 & \frac{e^{q_{2}}}{1 + e^{q_{1}} + e^{q_{2}}} & 0\\
          0 & \frac{1}{1 + e^{q_{3}} + e^{q_{4}}} & \frac{e^{q_{3}}}{1 + e^{q_{3}} + e^{q_{4}}} & \frac{e^{q_{4}}}{1 + e^{q_{3}} + e^{q_{4}}} & 0\\
          \frac{e^{q_{5}}}{1 + e^{q_{5}} + e^{q_{6}} + e^{q_{7}}} & \frac{e^{q_{6}}}{1 + e^{q_{5}} + e^{q_{6}} + e^{q_{7}}} & \frac{1}{1 + e^{q_{5}} + e^{q_{6}} + e^{q_{7}}} & \frac{e^{q_{7}}}{1 + e^{q_{5}} + e^{q_{6}} + e^{q_{7}}} & 0\\
          0 & \frac{e^{q_{8}}}{1 + e^{q_{8}} + e^{q_{9}}} & 0 & \frac{1}{1 + e^{q_{8}} + e^{q_{9}}} & \frac{e^{q_{9}}}{1 + e^{q_{8}} + e^{q_{9}}}\\
          \frac{e^{q_{10}}}{1 + e^{q_{10}} + e^{q_{11}} + e^{q_{12}}} & \frac{e^{q_{11}}}{1 + e^{q_{10}} + e^{q_{11}} + e^{q_{12}}} & 0 & \frac{e^{q_{12}}}{1 + e^{q_{10}} + e^{q_{11}} + e^{q_{12}}} & \frac{1}{1 + e^{q_{10}} + e^{q_{11}} + e^{q_{12}}}),
\end{equation}
}%
where $q_{1}, q_2, \dots, q_{12}$ are linear functions of the form 
$q_j = \zeta_{0,j} + \zeta_{1,j} \cdot z_k$, for $j \in \{1,2,\hdots, 12\}$,
and $z_k$ is the number of RBCs \textit{ordered} at time point $k$. Note that the $r^{th}$ row and $s^{th}$ column of $\vb{P}$, denoted $\vb{P}_{r,s}$, is defined as $\vb{P}_{r,s} := P(\bb_k^{(i)} = s \mid \bb_{k-1}^{(i)} = r)$ for $k \in \{2,3,\hdots, n_i\}$ and $r,s \in \{1,2,\hdots, 5\}$. We distinguish the number of RBCs \textit{ordered} versus the number of RBCs \textit{administered} because in most cases, there can exist a 30 minute delay from when RBCs are ordered versus administered. Therefore, RBC \textit{order} times serve as the covariate in the state transition probability matrix because they mostly coincide with detectable time instances of physiological change. The prior mass function for the latent state sequence of subject $i$ is then given by
$$p\big(\bb_1^{(i)} = s_{i,1}, \hdots, \bb_{n_i}^{(i)} = s_{i,n_i} \mid \bpi, \{\zeta_{0,j}, \zeta_{1,j}\}_{j=1}^{12}\big) = \bpi_{s_{i,1}} \cdot \prod_{k=2}^{n_{i}}\vb{P}_{s_{i,k-1}, s_{i,k}},$$
where $\bpi := (\bpi_{1}, \dots, \bpi_{5})^{\top}$ is a $5\times 1$ vector corresponding to the discrete, initial state distribution of the Markov process, and $s_{i,1},\dots,s_{i,n_i} \in \{1,\dots,5\}$.

\subsubsection{Conditional Response Model}\label{chap3:subsubsec:resp}
The response, $\bY^{(i)}$, conditional on the latent states, $\bb^{(i)}$, is modeled as:
{
\small\begin{align*}
    &\by^{(i)}_1 \mid \bb_1^{(i)} = s_{i,1}, \balpha^{(i)}, \bomega,\bbeta, \bR, \bA_{s_{i,1}}  &&\sim \text{N}_4\qty(\bnu_1^{(i)},\; \bGamma_{s_{i,1}}) \stepcounter{equation}\tag{\theequation}\label{chap3:eq:arEq}\\
    &\by^{(i)}_k \mid \by^{(i)}_{k-1}, \{\bb_j^{(i)} = s_{i,j}\}_{j=1}^k, \balpha^{(i)}, \bomega,\bbeta, \bR, \bA_{s_{i,k}}  &&\sim \text{N}_4\qty(\bnu_{k}^{(i)} + \bA_{s_{i,k}}\cdot(\by^{(i)}_{k-1} - \bnu_{k-1}^{(i)}),\; \bR), 
\end{align*} 
}%
for $k\in\{2,3,\hdots, n_i\}$, where
{
\small\begin{align*}
    &\bnu_1^{(i)} = g(\balpha^{(i)}, \vb*{b}^{(i)}_1) + \bD_{\omega, 1}^{(i)}\bomega + \bX_1^{(i)} \bbeta,\\
    &\bnu_k^{(i)} = g(\balpha^{(i)}, \vb*{b}^{(i)}_1) + \qty[\sum_{j=2}^{k}\mathbf{1}\{\bb^{(i)}_{j}=2\}] \balpha^{(i)}_{\cdot, 2} + \dots + \qty[\sum_{j=2}^{k}\mathbf{1}\{\bb^{(i)}_{j}=5\}] \balpha^{(i)}_{\cdot, 5}+ \bD_{\omega, k}^{(i)} \bomega + \bX_k^{(i)}\bbeta,\\
    &\bD_{\omega, k}^{(i)} = {\spacingset{1}\mqty(\vb*{0} & \vb*{0} \\ d_{h,k,1}^{(i)} \hdots d_{h,k,n_{hr}}^{(i)} & \vb*{0} \\ \vb*{0} & d_{m,k,1}^{(i)} \hdots d_{m,k,n_{map}}^{(i)}\\  \vb*{0} & \vb*{0} )}, \quad \bX_k^{(i)} = x_{k}^{(i)} \cdot \bI_4.
    \stepcounter{equation}\tag{\theequation}\label{chap3:eq:vec_mean_nu}
\end{align*}
}%
Let $\balpha^{(i)}$ be defined as a $4\times5$ matrix of random effect coefficients where the rows correspond to the four responses. The first column $\balpha^{(i)}_{\cdot, 1}$ corresponds to the subject-specific mean response when stable (i.e., state 1), and columns $\balpha^{(i)}_{\cdot, 2}, \balpha^{(i)}_{\cdot, 3}, \balpha^{(i)}_{\cdot, 4},$ and $\balpha^{(i)}_{\cdot, 5}$ correspond to the expected change in response during each of the four non-stable states (i.e., states 2, 3, 4, and 5, respectively). The intuition for not having a slope coefficient for state 1 is because all responses should exhibit no trend when stable. The remaining terms of (\ref{chap3:eq:arEq}) and (\ref{chap3:eq:vec_mean_nu}) are defined as follows: $\bomega$ is a $(n_{hr}+n_{map}) \times 1$ vector of medication effects (the first $n_{hr} = 34$ are medications affecting heart rate, and the last $n_{map} = 50$ are medications affecting MAP); $d_{h,k,\cdot}^{(i)}$ and $d_{m,k,\cdot}^{(i)}$ are the subject-specific doses for medications affecting heart rate and MAP, respectively, at time $k$; $\bbeta$ is a $4\times1$ vector of coefficients defining the effects of the \textit{administered} RBC transfusions on the mean process; $x_{k}^{(i)}$ is the number of RBC transfusions \textit{administered} up to time $k$; $\vb{A}_1, \hdots, \vb{A}_5$ are $4\times 4$ matrices of state-specific autocorrelation coefficients (similar to the model definition in \cite{li2015hemodynamic}); $\bR$ is the error covariance matrix; and $g(\balpha^{(i)}, \bb^{(i)}_1)$ is a random effect intercept term accounting for physiological variation upon ICU admission (see Section \ref{chap3:subsec:initAdj}). Note that $\bD_{\omega, k}^{(i)}$ has a sparse structure because medications do \textit{not} affect hemoglobin or lactate. Lastly, as suggested in Figure \ref{chap3:fig:ar-hmm}, the mean for subject $i$ at time $k$ is dependent on $\{\bb_j^{(i)} = s_{i,j}\}_{j=1}^k$.

Next, we assume a stable AR response by defining $\vb{A}_j := \text{diag}\{a_{1,j},a_{2,j},a_{3,j},a_{4,j}\}$ with $a_{1,j},a_{2,j},a_{3,j},a_{4,j} \in [0,1]$ for $j \in \{1,\hdots,5\}$. A standard assumption/property for the unconditional covariance, $\bGamma_{s_{i,1}}$, is that it satisfies {\small$\bGamma_{s_{i,1}} = \bA_{s_{i,1}} \bGamma_{s_{i,1}} \bA_{s_{i,1}}^T + \bR$}. Moreover, given the defined structure of $\vb{A}_{s_{i,1}}$, the $j^{th}$ row and $l^{th}$ column of $\bGamma_{s_{i,1}}$ has the following form 
$\qty[\bGamma_{s_{i,1}}]_{j,l} = [\bR]_{j,l} / (1-a_{j, s_{i,1}}a_{l, s_{i,1}})$. 
The autocorrelation coefficient matrices are state-dependent because it is often the case that heart rhythm changes as physiological conditions change \citep{latif2023traumatic}. Consequently, we account for heart rhythm variation by via state-dependent autocorrelation.

\subsubsection{Joint Conditional Density} \label{chap3:subsubsec:like}
The joint conditional density for the data is given by
{
\small 
\begin{align*}
    &f\qty(\{\bY^{(i)}\}_{i=1}^N \mid \{\bb^{(i)}\}_{i=1}^N, \{\balpha^{(i)}\}_{i=1}^N, \bomega, \bbeta, \bA_1, \hdots, \bA_5, \bR)= \prod_{i=1}^N f\qty(\by^{(i)}_1 \mid \bb_1^{(i)}=s_{i,1}, \balpha^{(i)}, \bomega,\bbeta, \bA_{s_{i,1}}, \bR)\\
    &\hspace{1in}\times \prod_{k=2}^{n_{i}}f\qty(\by^{(i)}_k \mid \by^{(i)}_{k-1}, \{\bb_j^{(i)} = s_{i,j}\}_{j=1}^k, \balpha^{(i)}, \bomega, \bbeta, \bA_{s_{i,k}}, \bR).
\end{align*}
}%
It is important to highlight that the parametric definition of the density is conditional on the unobserved states, $\{\bb^{(i)}\}_{i=1}^N$. Unlike traditional HMMs, the definition of our RSM makes it infeasible to marginalize/integrate over the latent state sequences. If it were feasible to integrate out the state sequences, then standard maximum-likelihood approaches exist for efficiently learning the model parameters of HMMs, and models alike. To handle our extra model complexity, we use a Bayesian computational approach, as outlined in Section \ref{chap3:subsec:bayesComp}, to learn both the posterior distribution of the response model parameters, as well as the discrete posterior distribution of the latent states. 

\subsubsection{Accounting for Physiological Changes Before ICU Admission}\label{chap3:subsec:initAdj}
It is unlikely that every patient admitted into the ICU will arrive in a stable state. More often, a patient has experienced some sort of trauma or stress prior to being admitted into the ICU. In order to account for this phenomenon, for a given subject $i$, the mean of the conditional response model in (\ref{chap3:eq:arEq}) is a function of $g(\balpha^{(i)}, \bb^{(i)}_1)$, which is defined as
\begin{equation}\label{chap3:eq:gDef}
    g(\balpha^{(i)}, \bb^{(i)}_1) := \balpha^{(i)}_{\cdot,1} + t_2^{(i)} \cdot \balpha^{(i)}_{\cdot,2} + \dots + t_5^{(i)} \cdot \balpha^{(i)}_{\cdot,5},
\end{equation}
where $t_2^{(i)}, t_3^{(i)}, t_4^{(i)}$, and $t_5^{(i)}$ represent the time spent in states 2, 3, 4, and 5, respectively, up to the initial ICU observation. Because no data exists before a subject's ICU admission, estimating $g(\balpha^{(i)}, \bb^{(i)}_1)$ is not feasible. However, not accounting for this added variability in the initial observation can lead to challenges in the estimation of other model parameters. One could naively assume either $t_2^{(i)} = t_3^{(i)} = t_4^{(i)} = t_5^{(i)} = 0$, $\forall i$, or 
$$g(\balpha^{(i)}, \bb^{(i)}_1) := \balpha^{(i)}_{\cdot,1} + \mathbf{1}\{\bb^{(i)}_{1}=2\} \cdot \balpha^{(i)}_{\cdot,2} + \dots + \mathbf{1}\{\bb^{(i)}_{1}=5\} \cdot \balpha^{(i)}_{\cdot,5}.$$
Both of these simplifications, however, would conflate the variability from the random effect, $\balpha^{(i)}$, with the variability due to non-zero $t_2^{(i)}, \dots, t_5^{(i)}$.

We can instead \textit{approximate} the conditional density for the data in the following manner. Given the initial observation, $\by^{(i)}_1$, and model parameters, $\bomega$ and $\bbeta$, we can approximate  
$g(\balpha^{(i)}, \bb^{(i)}_1) \approx \by^{(i)}_1 - \bD_{\omega, 1}^{(i)} \cdot \bomega - \bX_1^{(i)} \cdot \bbeta.$
Hence, define the approximate conditional model:
{
\small\begin{align*}
    & \bgamma^{(i)} \mid \by^{(i)}_{1}, \bomega, \bbeta, \vb*{G} && \sim \text{N}_4(\by^{(i)}_1 - \bD_{\omega, 1}^{(i)} \cdot \bomega - \bX_1^{(i)}\cdot \bbeta,\; \vb*{G}) \\
    &\by^{(i)}_1 \mid \bgamma^{(i)}, \bomega, \bbeta, \bR, \bA_{s_{i,1}}  &&\sim \text{N}_4\qty(\bnu_1^{(i)}, \bGamma_{s_{i,1}}) \stepcounter{equation}\tag{\theequation}\label{chap3:eq:arEq2}\\
    &\by^{(i)}_k \mid \by^{(i)}_{k-1}, \{\bb_j^{(i)} = s_{i,j}\}_{j=2}^k, \balpha_*^{(i)}, \bgamma^{(i)}, \bomega, \bbeta, \bR, \bA_{s_{i,k}}  &&\sim \text{N}_4\qty(\bnu_{k}^{(i)} + \bA_{s_{i,k}}\cdot(\by^{(i)}_{k-1} - \bnu_{k-1}^{(i)}), \bR), 
\end{align*} 
}%
for $k\in\{2,3,\hdots, n_i\}$, where $\vb*{G}$ is a new covariance matrix to learn, 
{
\small\begin{align*}
    \bnu_1^{(i)} &= \bgamma^{(i)} + \bD_{\omega, 1}^{(i)} \cdot \bomega + \bX_1^{(i)} \cdot \bbeta,\\
    \bnu_k^{(i)} &= \bgamma^{(i)} + \bD_{\alpha, k}^{(i)} \cdot \Vec\qty(\balpha^{(i)}_{*}) + \bD_{\omega, k}^{(i)} \cdot \bomega + \bX_k^{(i)} \cdot \bbeta,\\
    \bD_{\alpha, k}^{(i)} &= \bI_4 \otimes \mqty(\displaystyle\sum_{j=2}^{k}\mathbf{1}\{\bb^{(i)}_{j}=2\}& \displaystyle\sum_{j=2}^{k}\mathbf{1}\{\bb^{(i)}_{j}=3\} & \displaystyle\sum_{j=2}^{k}\mathbf{1}\{\bb^{(i)}_{j}=4\} & \displaystyle\sum_{j=2}^{k}\mathbf{1}\{\bb^{(i)}_{j}=5\}),
\end{align*}}%
$\Vec(\cdot)$ denotes the operation of forming a column vector from a matrix by stacking the columns of the matrix from left to right, and $\balpha^{(i)}_*$ is equal to $\balpha^{(i)}$ with the first column removed and then transposed (i.e., the four rows of $\balpha^{(i)}_*$ correspond to the four non-stable states and the four columns correspond to the four responses). Based on Bayesian computation routines, we can sample $\bgamma^{(i)}$ to then compute the conditional density. 

To further justify this approach, Supplementary Materials Section \ref{chap3:app:simpleSim} offers an additional, simpler simulation study. The purpose of this simulation is to illustrate that although (\ref{chap3:eq:arEq2}) is an \textit{approximate} conditional density, it is still effective at learning parameter estimates and recovering the underlying state sequences. This additional simulation also showcases the inferential consequences of \textit{not} accounting for pre-ICU-admission physiological changes. 

\subsection{Bayesian Computation}\label{chap3:subsec:bayesComp}
With the updated response model in (\ref{chap3:eq:arEq2}), the approximate joint posterior distribution is
{
\small\begin{align*}
    &\pi \qty(\{\bb^{(i)}\}_{i=1}^N, \{\balpha_*^{(i)}\}_{i=1}^N, \{\bgamma^{(i)}\}_{i=1}^N, \bomega, \bbeta, \bA_1, \hdots, \bA_5, \bR, \vb*{G}, \widetilde{\balpha}_*, \bUpsilon_\alpha, \bzeta, \bpi \;\Big |\; \{\bY^{(i)}\}_{i=1}^N)\\
    &\hspace{0.2in} \propto \Big\{ \prod_{i=1}^N \pi\qty(\balpha_*^{(i)} \mid \widetilde{\balpha}_*, \bUpsilon_\alpha) \cdot \pi\qty(\bgamma^{(i)} \mid \by^{(i)}_1, \bomega, \bbeta, \vb*{G}) \cdot \bpi_{s_{i,1}} \cdot f\qty(\by^{(i)}_1 \mid \bgamma^{(i)},\bomega, \bbeta, \bA_{s_{i,1}}, \bR)\\
    &\hspace{0.6in} \times \prod_{k=2}^{n_{i}}\vb{P}_{s_{i,k-1},s_{i,k}} \cdot f\qty(\by^{(i)}_k \mid \by^{(i)}_{k-1}, \{\bb_j^{(i)} = s_{i,j}\}_{j=1}^k, \balpha_*^{(i)}, \bgamma^{(i)}, \bomega, \bbeta, \bA_{s_{i,k}}, \bR) \Big \} \\
    &\hspace{0.6in} \times \pi(\bomega, \bbeta, \bA_1, \hdots, \bA_5, \bR, \vb*{G}, \widetilde{\balpha}_*, \bUpsilon_\alpha, \bzeta),
\end{align*}
}%
where the random effect and selected prior distributions are
\begin{align*}
    &\Vec(\balpha_*^{(i)}) \sim \text{N}_{16}(\Vec(\Tilde{\balpha}_*),\; \bUpsilon_\alpha), && \Vec(\Tilde{\balpha}_*) \sim \text{N}_{16}(\Vec(\Tilde{\balpha}_0),\; \bSigma_\alpha),\\
    &\bomega \sim \text{N}_{n_{hr}+n_{map}}(\bomega_0,\; \bSigma_\omega), && \bUpsilon_\alpha \sim \text{InvWish}(\vb*{\Psi}_\alpha,\; \nu_\alpha),\\
    &\bbeta \sim \text{N}_{4}(\bbeta_0,\; \bSigma_\beta), && \vb*{G} \sim \text{InvWish}(\vb*{\Psi}_G,\; \nu_G),\\
    &\bR \sim \text{InvWish}(\vb*{\Psi}_R,\; \nu_R).
\end{align*}
Gaussian/multivariate Gaussian priors are placed on the parameters or transformations of the parameters that are not displayed here. Parameter estimation is done using a Metropolis-within-Gibbs MCMC sampling algorithm; derivations and details are provided in Supplementary Materials Section \ref{chap3:app:bayes}. 

Precise prior specifications are found in Supplementary Materials Section \ref{chap3:app:priorSpec}, yet the following is a justification for some of the stronger priors. Notably, a diffuse prior on the noise parameter $\bR$ leads to a level of noise that inhibits the MCMC algorithm from accepting any states other than state 1. The intuition is that for a sufficiently large error variance, any change in the vitals is attributed to random chance rather than a physiological change. Consequently, we set the Inverse-Wishart prior degrees of freedom for $\bR$ to $\nu_R = 2 \cdot \sum_{i = 1}^N n_i$ and the prior scale matrix to $\bPsi_R = \nu_R \cdot \text{diag}\{\frac{1}{2}, \frac{3}{2}, \frac{3}{2}, \frac{1}{2}\}$. Next, $\bomega$ quantifies the medication effects; in particular, these medications are characterized as ``uppers'' or ``downers'' depending on their effects on heart rate and MAP. Hence, the specified prior on $\bomega$ encourages the medication effects for ``uppers'' to be positive and ``downers'' to be negative. The medication effects for Norepinephrine and Dexmedetomidine, in particular, necessitate stronger priors to estimate the correct sign.

Recall from Section \ref{chap3:subsec:initAdj} that the approximate conditional density is a function of the latent states and the random effects. Computing this approximate density is feasible because at every iteration of the MCMC, we sample from the conditional posterior distributions of $\{\bb^{(i)}\}_{i=1}^N$, $\{\balpha_*^{(i)}\}_{i=1}^N$, and $\{\bgamma^{(i)}\}_{i=1}^N$, respectively. For the purposes of sampling the random effect $\{\balpha_*^{(i)}\}_{i=1}^N$, we implement a rejection sampling strategy. In particular, to better classify the canonical characteristics of states 2 and 3, we sample each $\balpha_*^{(i)}$ according to its Gibbs update, but only ``accept'' if the proposed slopes for states 2 and 3 coincide with the expected hemodynamics with respect to the values being positive or negative. Lastly, for the Metropolis-Hastings (MH) updates, an adaptive proposal strategy is implemented during the burnin phase to improve the mixing/efficiency of the sampling by learning the proposal correlation structures and tuning the proposal variances/scales.

\subsection{State-Sampling Algorithm} \label{chap3:subsec:stateSamp}
The use of Bayesian estimation methods in state-space models to learn the posterior distribution of the latent state sequences has been studied in the literature before \citep[e.g.,][]{carter1994gibbs, djuric2002mcmc, scott2002bayesian, turek2016efficient, triantafyllopoulos2021bayesian, williams2024bayesian}. A detailed review of the various MCMC sampling approaches for the latent state process of an HMM is found in \cite{fearnhead2025mcmcstatespacemodels}. In our case, however (i.e., beyond the scope of HMMs or AR-HMMs), we have the additional constraint that the response model for a given time, $k$, is dependent on \textit{all} latent states up to time $k$. As such, the complex dependence structure coupled with the high-dimensional parameter space necessitates a novel and clever state-sampling procedure beyond a naive Gibbs or MH update; otherwise, learning the discrete posterior distribution of the latent state sequences quickly becomes computationally prohibitive. 

\subsubsection{Existing State-Sampling Methods} \label{chap3:subsubsec:existingSamp}

Let $p(\vb*{b}^{(i)}_k = s_{i,k}, \hdots, \vb*{b}^{(i)}_{k+p-1} = s_{i,k+p-1}  \mid \vb*{b}^{(i)}_{k-1} = s_{i,k-1}, \vb*{b}^{(i)}_{k+p} = s_{i,k+p},\; \dots)$ denote the full conditional distribution of the latent state sequence $\vb*{b}^{(i)}_k, \hdots, \vb*{b}^{(i)}_{k+p-1}$ for subject $i$ at starting time $k$, where $p \in \mathbb{Z}^+$. This is a discrete probability mass function where the realizations of $\vb*{b}_k^{(i)}, \hdots, \vb*{b}_{k+p-1}^{(i)}$ are dependent on the values of $\vb*{b}^{(i)}_{k-1}$ and $\vb*{b}^{(i)}_{k+p}$. With the restrictions on the allowable state transitions, 
let $\mathcal{B}^{(p)}_{s_{i,k-1}, s_{i,k+p}}$ denote the set of possible state $p$-tuples for $\vb*{b}^{(i)}_k, \hdots, \vb*{b}^{(i)}_{k+p-1}$ given $\vb*{b}^{(i)}_{k-1} = s_{i,k-1}$ and $\vb*{b}^{(i)}_{k+p} = s_{i,k+p}$ (e.g., $\mathcal{B}^{(1)}_{1,3} = \{\{2\}\}$ and $\mathcal{B}^{(2)}_{1,3} = \{\{1,2\}, \{2,2\}, \{2,3\}, \{4,2\}\}$). In the special edge cases of $k=1$ or $k = n_i - p + 1$, let $\mathcal{B}^{(p)}_{\boldsymbol{\cdot}, s_{i,k+p}}$ and $\mathcal{B}^{(p)}_{s_{i,k-1},\boldsymbol{\cdot}}$ define the set of possible state $p$-tuples, respectively.

Next, the state proposal distribution is derived by assigning probabilities to each element of $\mathcal{B}^{(p)}_{s_{i,k-1}, s_{i,k+p}}$. There exists a variety of approaches for defining the state proposal distribution in the literature; we will focus on two examples. The simplest approach is to assign equal probability to each element of $\mathcal{B}^{(p)}_{s_{i,k-1}, s_{i,k+p}}$ leading to an MH sampling procedure. This approach is seen in \cite{williams2024bayesian} and will be termed \textit{approach (A)}. Alternatively, in a Gibbs update, the proposal distribution can incorporate information from the data as well as the current response parameter values in the MCMC iteration. This state-sampling approach is seen in \cite{albert1993bayes, robert1993bayesian, robert1998reparameterization}, and will be termed \textit{approach (B)}. Their precise details are provided in Supplementary Materials Section \ref{chap3:app:altSamp}.

While approaches (A) and (B) are mathematically different proposal strategies, they both require computing $\mathcal{B}^{(p)}_{s_{i,k-1}, s_{i,k+p}}$ for any combination of $s_{i,k-1}, s_{i,k+p} \in \{1,\hdots, 5\}$. Additionally, a value for $p$ must be selected which, as we will demonstrate, can lead to computational challenges. In the case of approach (B), the Gibbs updates require looping over all elements of $\mathcal{B}^{(p)}_{s_{i,k-1}, s_{i,k+p}}$. As Table \ref{chap3:tab:sizeOfB} shows, an increase in $p$ leads to an exponential increase in the cardinality of $\mathcal{B}^{(p)}_{s_{i,k-1}, s_{i,k+p}}$ and consequently an exponential increase in compute time. Approach (A) does not exhibit the same computational inefficiency; however, it is subject to a statistical/algorithmic inefficiency in the sense that as the cardinality of $\mathcal{B}^{(p)}_{s_{i,k-1}, s_{i,k+p}}$ increases, the probability of randomly selecting a ``good'' proposal sequence decreases because each element in the set is assigned equal probability. In other words, each iteration of approach (B) is slower, but potentially many more iterations of approach (A) may be required. A more nuanced state-sampling strategy is needed.
{
\spacingset{1}
\begin{table}[!htb]
    \centering\footnotesize
    \begin{tabular}{|c|r r r r r r|}
    \hline
     & $p=1$ & $p=2$ & $p=4$ & $p=6$ & $p=8$ & $p=10$ \\
     \hline
     $|\mathcal{B}^{(p)}_{1,1}|$ & 1 & 3  & 37  & 395  & 4221  & 45123  \\
     $|\mathcal{B}^{(p)}_{2,3}|$ & 2 & 5  & 48  & 509  & 5436  & 58105  \\
     $|\mathcal{B}^{(p)}_{5,4}|$ & 4 & 13 & 137 & 1465 & 15661 & 167413 \\
     \hline
    \end{tabular}
    \caption{\footnotesize Examples of how the cardinality of $\mathcal{B}^{(p)}_{s_{i,k-1}, s_{i,k+p}}$ changes as a function of $p$.}
    \label{chap3:tab:sizeOfB}
\end{table}
}

\subsubsection{Our State-Sampling Procedure} \label{chap3:subsubsec:propDist}
Our novel state-sampling strategy can be found in Algorithm \ref{chap3:alg:statesamp} from Supplementary Materials Section \ref{chap3:app:stateSampAlgs}. To contextualize this algorithm, the state-sampling works as follows. The algorithm is an MH sampling routine. There exists flexibility in the number of time points, $p$, over which the state-space is sampled. In one extreme, for each subject, $i$, we can take $p = n_{i}$ and propose full state sequences (special case Algorithm \ref{chap3:alg:statesamp2}), or we can propose states over blocks of time instances in sequence with minimum $p=2$ (special case Algorithm \ref{chap3:alg:statesamp3}). 

For each step of the MCMC sampling routine, we sample $p \sim \text{uniform}\{2,\dots, 50\}$.  For a given subject, $i$, and starting time, $k \in \{1,\hdots, n_i-2\}$, Algorithm \ref{chap3:alg:statesamp} makes state proposals in accordance to the likelihood and transition probabilities across time points $\{k, k+1, \hdots, k_{\text{max}}\}$ where $k_{\text{max}} = \min\{k+p-1, n_i\}$. For most values of $p$ and $k$, the discrete proposal distribution for $\{\bb_j^{(i)} = s_{i,j}\}_{j=k}^{k_{\text{max}}}$ is defined as 
{
\footnotesize
\[
\begin{split}
    &q\qty(\{\vb*{b}_j^{(i)} = s_{i,j}\}_{j=k}^{k_{\text{max}}} \mid \{\vb*{b}_j^{(i)} = s_{i,j}\}_{j\notin \{k,\dots, k_{\text{max}}\}}, \vb*{Y}^{(i)}, \balpha_*^{(i)}, \bgamma^{(i)}, \bomega, \bbeta, \bA_1, \hdots, \bA_5, \bR,\bzeta,\bpi)\\
    &\hspace{0.15in}:=\begin{cases}
        \frac{\bpi_{s_{i,1}} \cdot f\qty(\vb*{y}^{(i)}_1 \mid \vb*{b}_1^{(i)}=s_{i,1},\; \text{rest})}{\sum_{m=1}^5 \bpi_{m} \cdot f\qty(\vb*{y}^{(i)}_1 \mid \vb*{b}_1^{(i)}=m, \; \text{rest})}
        {\displaystyle\prod_{t=k+1}^{k_{\text{max}}-2}} \frac{\vb{P}_{s_{i,t-1},s_{i,t}} \cdot f\qty(\vb*{y}^{(i)}_t \mid \vb*{y}^{(i)}_{t-1}, \{\vb*{b}_j^{(i)} = s_{i,j}\}_{j=1}^t, \; \text{rest})}{\sum_{m=1}^5\vb{P}_{s_{i,t-1},m} \cdot f\qty(\vb*{y}^{(i)}_t \mid \vb*{y}^{(i)}_{t-1}, \{\vb*{b}_j^{(i)} = s_{i,j}\}_{j=1}^{t-1}, \vb*{b}_t^{(i)} = m, \; \text{rest})}\\
        \hspace{0.5in}\times \qty|\mathcal{B}^{(2)}_{s_{i,k_{\text{max}}-2}, s_{i,k_{\text{max}}+1}}|^{-1} &,\; k = 1\\
        {\displaystyle\prod_{t=k}^{k_{\text{max}}-2}} \frac{\vb{P}_{s_{i,t-1},s_{i,t}} \cdot f\qty(\vb*{y}^{(i)}_t \mid \vb*{y}^{(i)}_{t-1}, \{\vb*{b}_j^{(i)} = s_{i,j}\}_{j=1}^t, \; \text{rest})}{\sum_{m=1}^5\vb{P}_{s_{i,t-1},m} \cdot f\qty(\vb*{y}^{(i)}_t \mid \vb*{y}^{(i)}_{t-1}, \{\vb*{b}_j^{(i)} = s_{i,j}\}_{j=1}^{t-1}, \vb*{b}_t^{(i)} = m, \; \text{rest})}\cdot \qty|\mathcal{B}^{(2)}_{s_{i,k_{\text{max}}-2}, s_{i,k_{\text{max}}+1}}|^{-1} & \text{, else}
    \end{cases}.
\end{split} \stepcounter{equation}\tag{\theequation}\label{chap3:eq:propDist}
\]
}%
See Supplementary Materials Section \ref{chap3:app:stateSampInfo} for other/all cases.

Similar to approach (B), Algorithm \ref{chap3:alg:statesamp} proposes state sequences in proportion to the likelihood given the model parameters of the current MCMC iteration. Whereas approach (B) has computational complexity $\mathcal{O}(5^p)$, our Algorithm \ref{chap3:alg:statesamp} achieves complexity $\mathcal{O}(5(p-2) + 5^2)$. Algorithm \ref{chap3:alg:statesamp} shares a similar structure to the ``forward-backward Gibbs sampler'' seen in \cite{scott2002bayesian}. 

To better illustrate the efficiency of Algorithm \ref{chap3:alg:statesamp}, Table \ref{chap3:tab:sampComputeTime} presents the median compute time for one iteration of an MCMC sampling routine using Algorithm \ref{chap3:alg:statesamp} versus approaches (A) and (B) on a dataset with 10 simulated subjects. As a baseline, we initialize each algorithm with the state sequence having all states being state 1. For approach (A), the accuracy is maximized when $p=4$, and decreases for larger $p$. This highlights the intuition given in Section \ref{chap3:subsubsec:existingSamp} that as the cardinality of $\mathcal{B}^{(p)}_{s_{i,k-1}, s_{i,k+p}}$ increases, the true state sequence has a lower chance of being proposed via uniform sampling at each iteration, and so more iterations of the algorithm are needed to happen upon it. Next, we see that approach (B) with $p=8$ leads to the second highest accuracy, though, it only completed 36 out of the 100 iterations in 12 hours. When compared to Algorithm \ref{chap3:alg:statesamp} [which obtained the highest accuracy], approach (B) with $p=8$ is 36,394 times slower. After the 100 iterations, Algorithm \ref{chap3:alg:statesamp} achieves the fastest median compute time \textit{and} the highest state accuracy.
{
\spacingset{1}
\begin{table}[!htb]
    \centering\footnotesize
    \begin{tabular}{|l|l|l|}
        \hline
        Approach & Time (sec.) & Percent correct \\
        \hline
        A ($p = 2$)  & 0.5598 & 0.6340\\
        A ($p = 4$)  & 0.8022 & 0.6481\\
        A ($p = 6$)  & 0.8813 & 0.6114\\
        A ($p = 8$)  & 1.0659 & 0.5472\\
        A ($p = 10$) & 4.5732 & 0.5240\\
        \hline
        B ($p = 2$)  & 1.9612 & 0.6939\\
        B ($p = 4$)  & 24.966 & 0.7673\\
        B ($p = 6$)  & 267.83 & 0.8082\\
        B ($p = 8$)  & $2653.1^*$ & $0.8392^*$\\
        B ($p = 10$) & $29333.9^*$ & $0.8258^*$\\
        \hline
        Algorithm \ref{chap3:alg:statesamp} & 0.0729 & 0.8766\\
        \hline
    \end{tabular}
    \caption{\footnotesize Median compute times (in seconds) over 100 iterations of each state-sampling algorithm and the accuracy of each sampling algorithm after the 100 iterations.  Accuracy is defined as the proportion of time instances where the posterior modal state (after 100 runs) correctly corresponds to the true state. The $(*)$ indicates that the 100 iterations did not finish within 12 hours. In particular, the Gibbs update (approach (B)) with $p=10$ only completed \textit{three} iterations before the algorithm timed out. Note that Algorithm \ref{chap3:alg:statesamp} does not change with respect to $p$ because at each iteration of the MCMC routine, a new $p$ is randomly selected from $\{2,\hdots, 50\}$.}
    \label{chap3:tab:sampComputeTime}
\end{table}
}

Lastly, recall from Section \ref{chap3:subsubsec:clinic} that we can use the number of RBC transfusions to serve as partial labels in our training data. In particular, if a patient receives three or more RBC transfusions in any 12 hour window, they have likely suffered some sort of hemorrhagic event. This indicator is built into our state-sampling in the following manner. For patient encounters satisfying this rule, when a state sequence is proposed, the sequence must contain state 2 at some point before the last of the RBC transfusion order times in the 12 hour window, else rejected. In addition to this heuristic, a small subset of the patient encounters have been evaluated by our three anesthesiologists to provide a yes/no answer as to whether the patient suffered from internal bleeding during their encounter. If the patient was clinically annotated ``yes,'' then proposed state sequences, again, must contain at least one state 2. Conversely, if the patient was clinically annotated ``no,'' then the only allowable states are 1, 4, and 5. Note, however, that these rules are removed for the case study in Section \ref{chap3:subsec:caseStudy}.

\section{Simulation Study}\label{chap3:sec:sim}

The lack of state labels in the real data necessitates a thorough simulation study to test our model's ability to detect internal bleeding. This simulation study serves a crucial role as it lays the foundation for how we interpret the discrete posterior probabilities of the physiological states. 
If we consider applying our approach in a hospital setting, we need to quantify what a ``high'' posterior probability of internal bleeding is such that we can adequately alarm for clinical intervention. Therefore, this simulation will determine a threshold/cutoff for the posterior probability of internal bleeding that is calibrated to balance the sensitivity of identifying state 2 with minimizing the number of false alarms. This threshold is then applied in our real case study in Section \ref{chap3:subsec:caseStudy}.

\subsection{Data Generating Mechanism}\label{chap3:simDataGen}
In order to simulate data that closely resembles the observed training data, we take the exact covariate information, length of patient encounters, and level of missingness from the real data. The true parameter values are determined from training the model on the real data as in Section \ref{chap3:sec:realDataAnalysis}; details are provided in Supplementary Materials Section \ref{chap3:app:simulationPar}. Given the covariate information and ``true'' parameter values, the latent state sequences and longitudinal response measurements are simulated as follows.

The data generating mechanism is identical to the model definition provided in Sections \ref{chap3:subsubsec:latent} and \ref{chap3:subsubsec:resp}, with one exception. First, for each subject $i \in \{1,\hdots, N\}$, the latent state sequence is simulated according to the transition probability matrix defined in (\ref{chap3:eq:transProb}). Recall from Section \ref{chap3:subsec:initAdj} that our model accounts for possible physiological changes prior to a patient's ICU admission. Thus, we need to simulate values for $t_2^{(i)}, \hdots, t_5^{(i)}$ from (\ref{chap3:eq:gDef}). With $n_i$ representing the length of the observed patient encounter, we generate a state sequence of length $n_i + m_i$, where $m_i \sim \text{uniform}\{0,1,\hdots, 50\}$. The first $m_i$ time points allow for possible state changes before the start of the patient encounter, with the initial state always being state 1. The first $m_i$ states are generated according to (\ref{chap3:eq:transProb}), except we assume $q_j = \zeta_{0,j}$, $\forall j \in \{1,\hdots, 12\}$. The last $n_i$ time points correspond to the observed process. Let $\vb*{b}_{long}^{(i)}$ denote the state sequence of length $n_i + m_i$, and $\vb*{b}^{(i)}$ represent the true state sequence for subject $i$ (i.e., the last $n_i$ states of $\vb*{b}_{long}^{(i)}$). Then $t_l^{(i)} := \sum_{k = 1}^{m_i+1} \mathbf{1}\{\vb*{b}_{long,\; k}^{(i)} = l\}$ for $l \in \{2,3,4,5\}$ is used to define $g(\balpha^{(i)}, \bb^{(i)}_1)$. Then, we generate the responses according to model (\ref{chap3:eq:arEq}), setting the proportion of missingness to that of the actual patient encounter.

One hundred datasets are simulated, with each containing 1,000 patient encounters (500 training and 500 testing), and the MCMC algorithm is run for 10,000 steps with the first 5,000 discarded as burnin.

\subsection{Interpretation of Results and Posterior Probabilities} \label{chap3:subsec:postInt}

One approach to evaluate model performance in identifying the latent state sequence for each subject is to compare the posterior modal state at each time point to the true state (as previously done in Table \ref{chap3:tab:sampComputeTime}). Using this metric, the median accuracy across all datasets is 0.7393. However, for our purposes, we are more interested in how precise our model is at detecting specifically state 2. Posterior probabilities are not necessarily calibrated to frequentist/aleatory probabilities, and so we must determine the minimum posterior probability of bleeding threshold $c \in [0,1]$ that best identifies the onset of state 2.

\begin{figure}[!htb]
    \centering
    \spacingset{1}
    \includegraphics[width=0.55\linewidth]{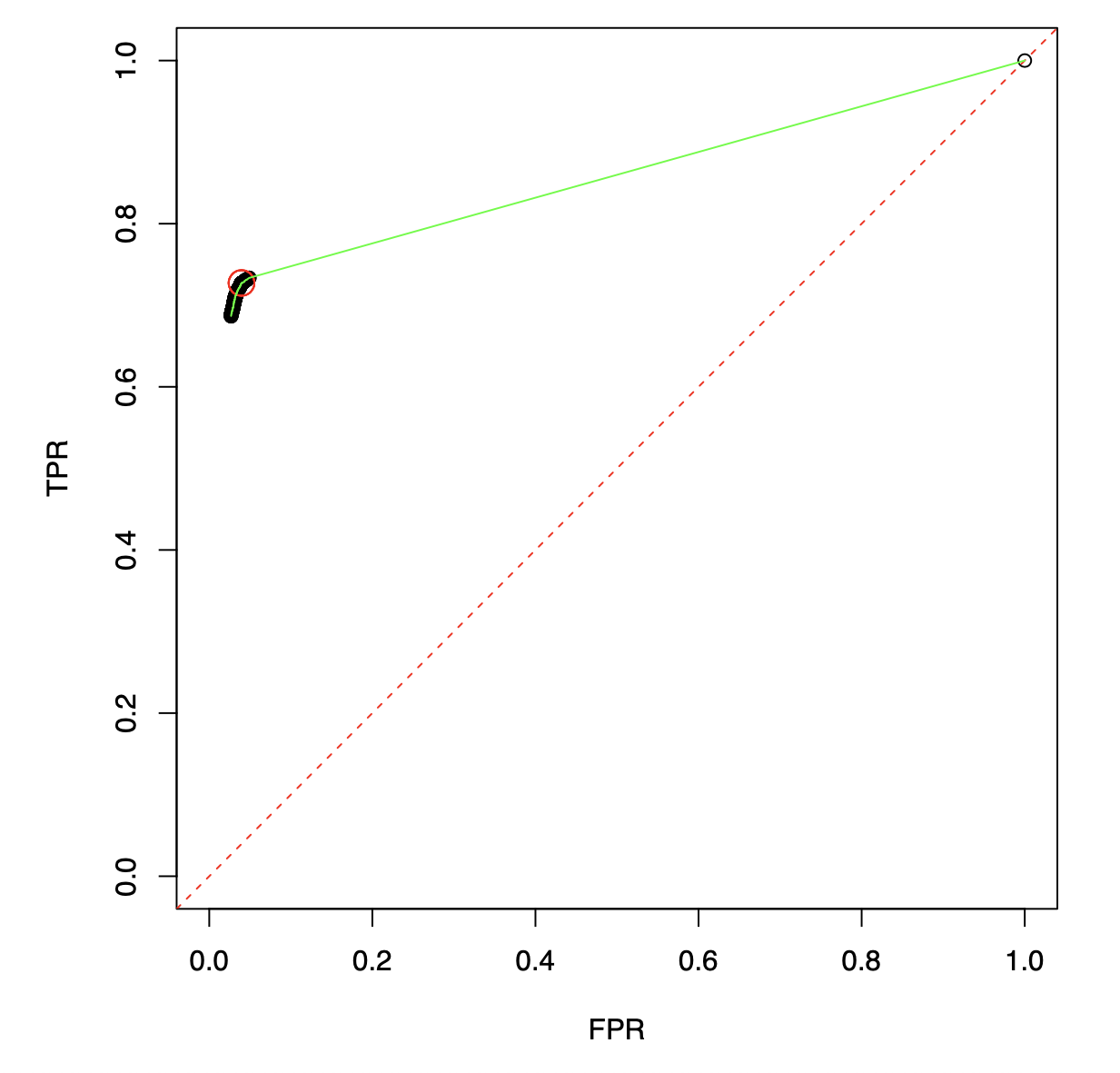}
    \caption{\footnotesize ROC curve depicting the sensitivity and specificity of identifying state 2. The red circle indicates the posterior probability threshold $c \in [0,1]$ that minimizes FPR - TPR.}
    \label{chap3:fig:ROC}
\end{figure}
Figure \ref{chap3:fig:ROC} presents the ROC curve for the identification of state 2 for \textit{one} of the 100 simulated datasets. For each simulated dataset, a value of $c$ is determined by finding the one that minimizes the difference between the false positive rate (FPR) and true positive rate (TPR). Using the median value of $c$ across all datasets as a threshold, we get $\hat{c} = 0.0465$. The median out-of-sample AUC (computed using the trapezoidal integral approximation) across all simulations is 0.8419. Using the threshold $\hat{c} = 0.0465$, the median sensitivity of correctly identifying state 2 is 0.7270, and the median specificity of correctly identifying \textit{not} state 2 is 0.9621. Table \ref{chap3:tab:fiveSummary} provides additional summary statistics. Trace plots and box plots for the simulation study are provided in Supplementary Materials Section \ref{chap3:app:simulationTrace}.

\begin{table}[!htb]
\spacingset{1}
    \centering\footnotesize
    \begin{tabular}{|l| c c c c|}
    \hline
    & Q1 & Median & Mean & Q3\\
    \hline
        threshold value $c$ & 0.0208 & 0.0465 & 0.0543 & 0.0700\\
        out-of-sample AUC & 0.8303 & 0.8419 & 0.8400 & 0.8501 \\
        sensitivity of state 2  & 0.7027 & 0.7270 & 0.7204 & 0.7412\\
        specificity of not state 2  & 0.9581 & 0.9621 & 0.9622 & 0.9655\\
        \hline
    \end{tabular}
    \caption{\footnotesize Summary statistics across 100 simulated datasets. The sensitivity and specificity calculations are done using $\hat{c} = 0.0465$.}
    \label{chap3:tab:fiveSummary}
\end{table}

An important takeaway from this simulation study is how small the threshold for the posterior probability of bleeding is. Recall the true parameter values for the simulation come from the trained model in Section \ref{chap3:sec:realDataAnalysis}. Based on the parameter estimates in Table \ref{chap3:tab:zetaEst}, the probability of transitioning out of states 1 or 5 is relatively \textit{low}, whereas the probability of transitioning out of states 2, 3, or 4 is \textit{high}. Namely, state 2 is a rare event meaning that it is indicated by even a small posterior probability in the fitted RSM.

\section{Real Data Analysis}\label{chap3:sec:realDataAnalysis}

\subsection{Results from Training Data} \label{chap3:subsec:realResults}
\begin{figure}
    \centering
    \spacingset{1}
    \includegraphics[page = 1, width=0.49\linewidth]{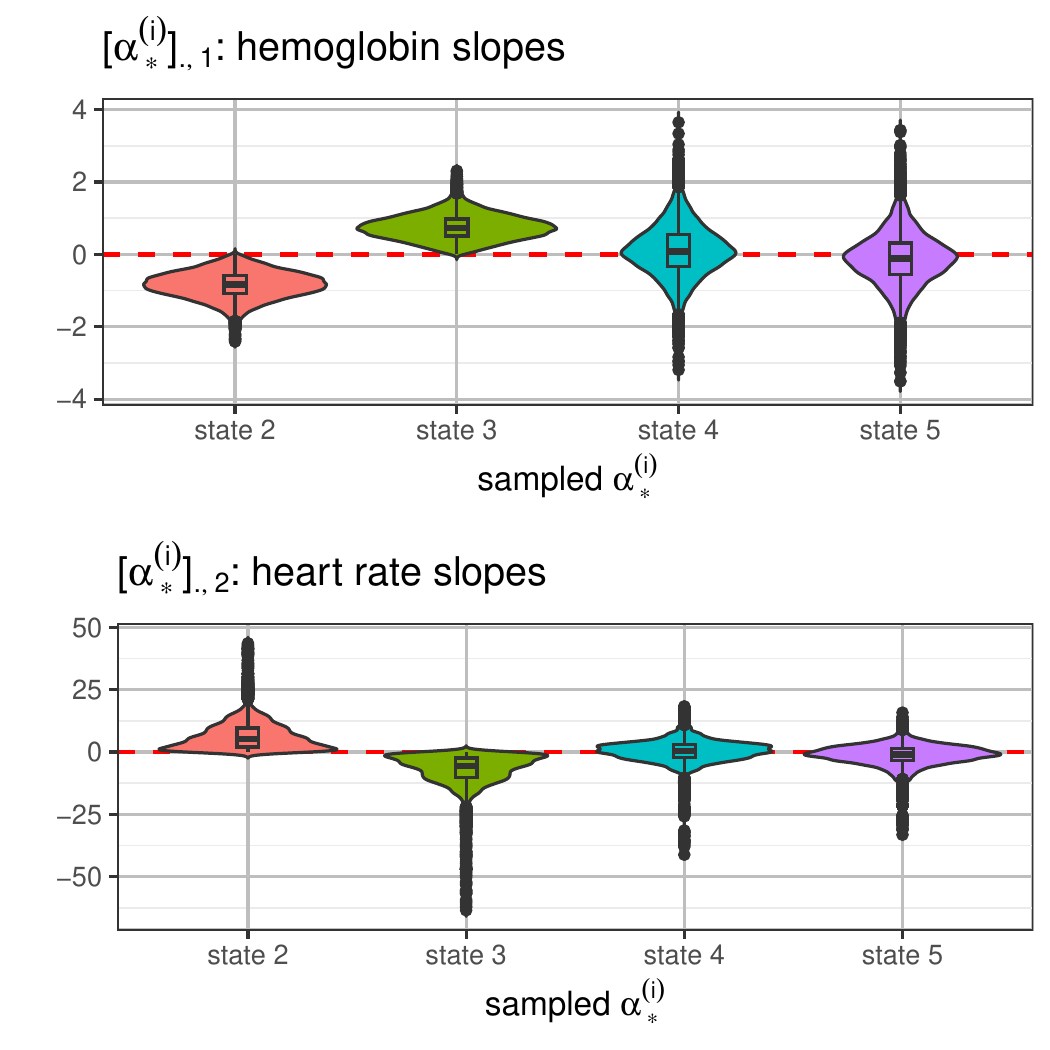}
    \includegraphics[page = 2, width=0.49\linewidth]{Plots/sampled_alpha_0_it4.pdf}
    \caption{\footnotesize Sampled random effect coefficients across all 500 training subjects for the last MCMC iteration.}
    \label{chap3:fig:sampAlpha}
\end{figure}

Recall from Section \ref{chap3:sec:method} that $\Tilde{\balpha}_*$ and $\bUpsilon_\alpha$ are the random effect mean and variance terms, respectively, which characterize the expected change in vital response based on the current and previous latent states. In addition, recall that $\bzeta$ are the transition rate coefficients characterizing the evolution of the latent state process, and $\{\bA_j\}_{j=1}^5$ are the state-dependent AR coefficient matrices quantifying the autocorrelation in response outcomes over time. The estimates of $\Tilde{\balpha}_*$, $\bUpsilon_\alpha$, $\bzeta$, and $\{\bA_j\}_{j=1}^5$ offer particularly useful insight into understanding the physiological characteristics of onset and offset of bleeding. Figure \ref{chap3:fig:sampAlpha} visualizes sampled random effect coefficients from the fitted RSM. Tables \ref{chap3:tab:zetaEst} and \ref{chap3:tab:arEst} provide the posterior median estimates and 95\% credible intervals for $\bzeta$ and $\{\bA_j\}_{j=1}^5$, respectively.

Figure \ref{chap3:fig:sampAlpha} and Table \ref{chap3:tab:zetaEst} characterize the expected phenotypes for the different latent states. Recall that during a bleeding event (state 2), it is expected that hemoglobin decreases, heart rate increases, MAP decreases, and lactate increases. The relation to zero of the violin plots in Figure \ref{chap3:fig:sampAlpha} are consistent with these behaviors. We must acknowledge that the rate of bleeding varies for every patient experiencing blood loss, and changes in vital signs and lab values are frequently delayed. Nonetheless, to summarize a typical presentation of a patient suffering from a clinically significant bleeding event, the fitted model suggests the following changes over a 15 minute period: a decrease in hemoglobin of 0.846 g/dL, an increase in heart rate of 6.559 bpm, a decrease in MAP of 9.623 mmHg, and an increase in lactate of 0.749 mmol/L. Figure \ref{chap3:fig:sampAlpha} also illustrates that the inter-individual variability in hemorrhage phenotype is quite large. For example, the violin plots for the slope coefficients on heart rate and MAP during bleeding suggest that many exhibit trends similar to the estimated parent mean $\Tilde{\balpha}_*$; however, there are also some subjects exhibiting changes in heart rate and MAP of over 40 units in magnitude in a 15 minute period. Additionally, for a given vital, we see that the sampled slope coefficients for state 3 are roughly a mirrored image of the sampled slopes for state 2, which aligns with the intuition that state 3 is characterizing the recovery from bleeding.

{
\spacingset{1}
\begin{table}[!htb]
    \centering
    \footnotesize
    \begin{tabular}{|l|l|l|}
    \hline
    \multicolumn{3}{|c|}{State Transitions} \\
    \hline
    \textbf{1 $\to$ 2} & \textbf{1 $\to$ 4} & \textbf{2 $\to$ 3} \\
    $\widehat{\zeta}_{0,1} = -2.996 [-3.069, -2.924]$ 
    & $\widehat{\zeta}_{0,2} = -2.127 [-2.174, -2.078]$ 
    & $\widehat{\zeta}_{0,3} = 1.94 [1.824, 2.055]$ \\
    $\widehat{\zeta}_{1,1} = 0.542 [0.172, 0.851]$ 
    & $\widehat{\zeta}_{1,2} = 0.231 [-0.188, 0.55]$ 
    & $\widehat{\zeta}_{1,3} = -0.513 [-1.064, 0.087]$\\
    \hline \hline
    \textbf{2 $\to$ 4} & \textbf{3 $\to$ 1} & \textbf{3 $\to$ 2} \\
    $\widehat{\zeta}_{0,4} = 0.616 [0.48, 0.754]$ 
    & $\widehat{\zeta}_{0,5} = 1.095 [0.979, 1.211]$ 
    & $\widehat{\zeta}_{0,6} = 0.145 [0.016, 0.274]$\\
    $\widehat{\zeta}_{1,4} = 0.105 [-0.449, 0.701]$ 
    & $\widehat{\zeta}_{1,5} = -0.763 [-1.793, 0.144]$ 
    & $\widehat{\zeta}_{1,6} = -0.136 [-0.989, 0.662]$\\
    \hline \hline
    \textbf{3 $\to$ 4} & \textbf{4 $\to$ 2} & \textbf{4 $\to$ 5} \\
    $\widehat{\zeta}_{0,7} = 0.301 [0.158, 0.438]$ 
    & $\widehat{\zeta}_{0,8} = -1.056 [-1.143, -0.97]$ 
    & $\widehat{\zeta}_{0,9} = 0.726 [0.674, 0.78]$\\
    $\widehat{\zeta}_{1,7} = -0.594 [-1.714, 0.348]$ 
    & $\widehat{\zeta}_{1,8} = 0.508 [-0.056, 1.095]$ 
    & $\widehat{\zeta}_{1,9} = 0.054 [-0.4, 0.681]$\\
    \hline \hline
    \textbf{5 $\to$ 1} & \textbf{5 $\to$ 2} & \textbf{5 $\to$ 4} \\
    $\widehat{\zeta}_{0,10} = -0.178 [-0.232, -0.121]$ 
    & $\widehat{\zeta}_{0,11} = -1.981 [-2.101, -1.871]$ 
    & $\widehat{\zeta}_{0,12} = -0.483 [-0.549, -0.419]$\\
    $\widehat{\zeta}_{1,10} = -0.449 [-1.008, 0.007]$ 
    & $\widehat{\zeta}_{1,11} = 0.156 [-0.367, 0.557]$ 
    & $\widehat{\zeta}_{1,12} = -0.026 [-0.444, 0.318]$\\
    \hline
    \end{tabular}
    \caption{\footnotesize The posterior median estimates and 95\% credible sets for the transition rate parameters.}
    \label{chap3:tab:zetaEst}
\end{table}
}%

In terms of states 4 and 5, namely NBE and NBER, we see more variation in the sampled random effect slopes in Figure \ref{chap3:fig:sampAlpha} as compared to states 2 and 3. It is difficult to precisely characterize a physiological interpretation for states 4 and 5 because while they can account for more variable hemodynamic changes in vitals, they also appear to be indicating more prolonged physiological conditions, with reference to the transition rate coefficient estimates in Table \ref{chap3:tab:zetaEst}. 

A basic reading of Table \ref{chap3:tab:zetaEst} indicates that transitioning from state 1 to either state 2 or 4 (in any 15 minute unit of time) is rarer than any other valid state transition. Furthermore, state 1 $\to$ 2 is \textit{less} likely than state 1 $\to$ 4, i.e., consistent with state 2 characterizing only internal bleeding, whereas state 4 is meant to characterize all other physiological changes. 

Lastly, Table \ref{chap3:tab:arEst} provides an additional perspective on how best to interpret the five states. Clinical intuition suggests that patients' vitals are more highly autocorrelated when stable, whereas they appear more sporadic and less autocorrelated in the presence of shock. As suggested in Table \ref{chap3:tab:arEst} as well as by our \textit{a priori} belief, heart rate, MAP, and lactate are highly autocorrelated when a patient is stable. However, for states 2, 3, 4, and 5, the autocorrelation between successive heart rate and MAP measurements is much less. Because all states aside from state 1 represent non-stable physiological conditions, the low degree of autocorrelation aligns with our previous intuition. As for the autocorrelation coefficients on hemoglobin and lactate, we do not see as drastic a change across the five states as compared with heart rate and MAP. This is likely a result confounded with the high degree of missingness in these two vitals.

{
\spacingset{1}
\begin{table}[!htb]
    \centering
    \footnotesize
    \begin{tabular}{|l|c|c|c|c|}
     \hline
     & hemoglobin & heart rate  & MAP & lactate \\
     \hline
     state 1 ($\bA_1$) & $0.142 [0.048, 0.265]$ 
     & $0.969 [0.966, 0.972]$ 
     & $0.880 [0.873, 0.887]$ 
     & $0.863 [0.838, 0.884]$\\
     state 2 ($\bA_2$) & $0.197 [0.062, 0.385]$ 
     & $0.004 [0.002, 0.008]$ 
     & $0.002 [0.001, 0.003]$ 
     & $0.775 [0.688, 0.848]$\\
     state 3 ($\bA_3$) & $0.494 [0.130, 0.824]$ 
     & $0.273 [0.217, 0.299]$ 
     & $0.009 [0.004, 0.018]$ 
     & $0.935 [0.863, 0.978]$\\
     state 4 ($\bA_4$) & $0.274 [0.085, 0.633]$ 
     & $0.001 [0.001, 0.003]$ 
     & $0.001 [0.001, 0.002]$ 
     & $0.772 [0.642, 0.868]$\\
     state 5 ($\bA_5$) & $0.101 [0.036, 0.198]$ 
     & $0.364 [0.339, 0.387]$ 
     & $0.143 [0.098, 0.171]$ 
     & $0.260 [0.122, 0.383]$\\
     \hline
    \end{tabular}
    \caption{\footnotesize The posterior median estimates and 95\% credible sets for the vector AR coefficients.}
    \label{chap3:tab:arEst}
\end{table}
}%

\subsection{Case Study} \label{chap3:subsec:caseStudy}
Using the parameter estimates from Section \ref{chap3:subsec:realResults}, we implement our state-sampling routine on a test set comprised of five real patient encounters (Subjects A, B, C, D, and E). The MCMC routine is run for 2,000,000 iterations where the only parameters being sampled are the random effect coefficients $\balpha_*^{(i)}$ and $\bgamma^{(i)}$, and the state sequences $\vb*{b}^{(i)}$. All remaining parameters are fixed at their posterior medians. Using the information from the last 500,000 steps, Figure \ref{chap3:fig:chartPlot3} presents the results for one out of the five test subjects (Subject C), and the remaining four test subject case studies are in Supplementary Materials Section \ref{chap3:app:chartTest}. 

\subsubsection{Subject C}
\begin{figure}[!htb]
    \centering
    \spacingset{1}
    \includegraphics[page = 3, width=\linewidth]{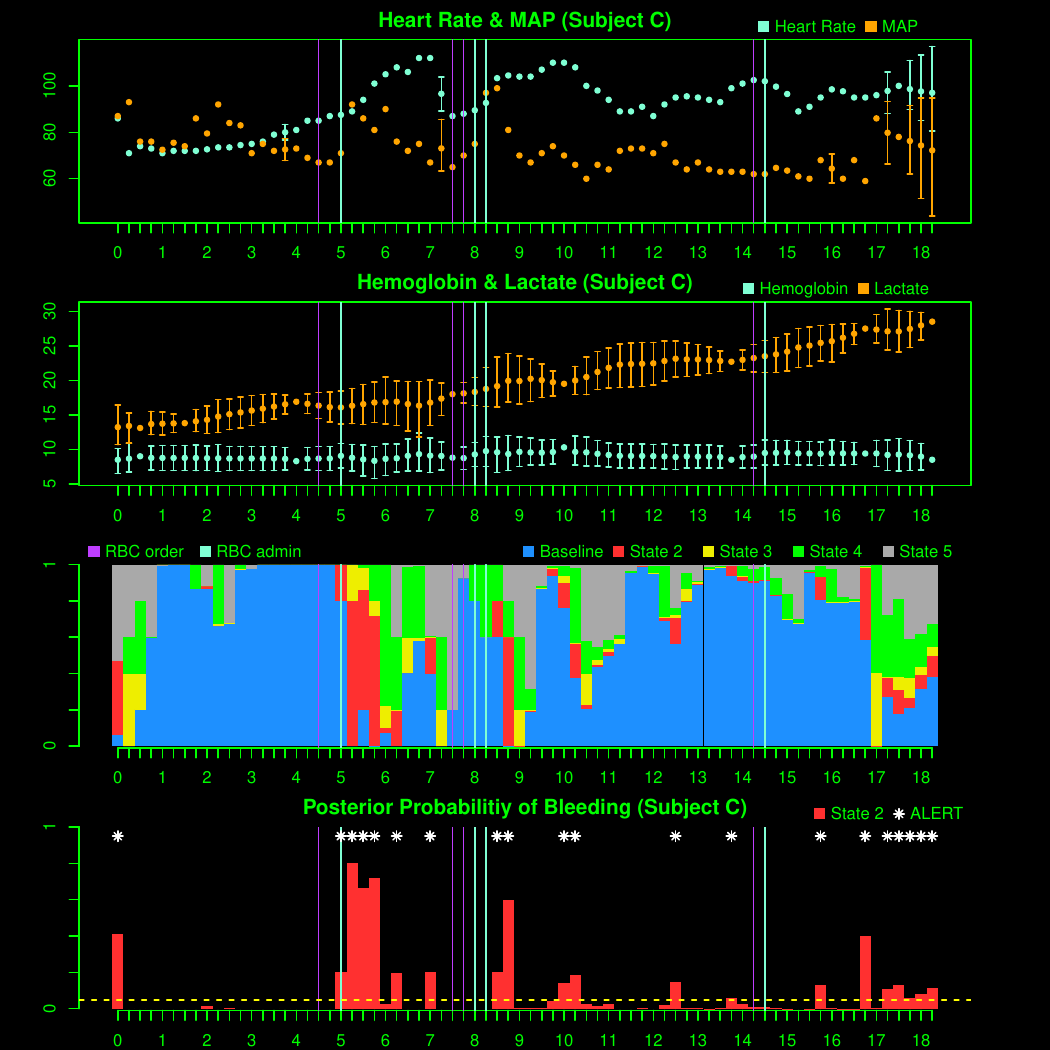}
    \caption{\footnotesize The top two panels correspond to the longitudinal vital measurements. The points with the error bars correspond to missing values; the error bars are empirical 95\% credible intervals for the imputed response values. The third panel depicts the discrete posterior probability distributions of the latent states, at each time point. The bottom panel is the posterior probability of state 2 at each time point, and the yellow dashed line represents the threshold $\hat{c} = 0.0465$ determined from Section \ref{chap3:subsec:postInt}. The white stars indicate that the posterior probability of state 2 exceeds the threshold. The purple and turquoise vertical lines represent RBC transfusion order and administration times, respectively.}
    \label{chap3:fig:chartPlot3}
\end{figure}

Figure \ref{chap3:fig:chartPlot3} follows a 72-year-old woman who underwent resection and replacement of a thoracoabdominal aortic aneurysm. Her operative course was notable for significant blood loss and coagulopathy requiring multiple transfusions.  During the first several hours of admission to the ICU post-operatively (shown in Figure \ref{chap3:fig:chartPlot3}), she had 2 episodes of increased chest tube output in the setting of coagulopathy (INR 2.4, fibrinogen 84, platelets 18) for which she received plasma, cryoprecipitate, platelets, and RBCs (only RBCs indicated in Figure \ref{chap3:fig:chartPlot3}). Over the next 12-16 hours, she developed refractory acidosis and progressive hemodynamic deterioration consistent with septic shock. A CT angiogram revealed a superior mesenteric artery (SMA) dissection, and she was rapidly taken back to the operating room. Unfortunately, half of her small bowel was found to be ischemic, and her colon was necrotic requiring a total colectomy. The SMA bypass graft was revised, and her abdomen was packed open. Blood loss was estimated at 350 mL. To optimize bowel perfusion, she was transfused with 2 additional units of RBCs. Despite aggressive care after returning to the ICU, her shock state progressed with severe metabolic acidosis, escalating vasopressor and inotropic support and eventual myocardial dysfunction (an overall state of mixed shock including septic, hypovolemic due to capillary leak, and cardiogenic). Care was withdrawn.

Figure \ref{chap3:fig:chartPlot3} appears to reflect the patient’s dynamic course with temporal accuracy as reflected by the transition in states. Although the figure indicates state 2 suggesting bleeding on several occasions (as seen in red), these events were not always clinically correlated with hemorrhage. Instead, state 2 was suggested when septic and cardiogenic shock were present and progressing rapidly. This is likely due to the overlap in the presentation of different shock types in terms of the vital signs heart rate, MAP, and lactate.

\section{Conclusion}\label{chap3:sec:conc}

The successful development of the RSM described in this manuscript will advance patient monitoring and medical interventions, particularly for transfusion medicine. Early detection of occult bleeding has been a persistent and difficult challenge for clinicians. The expected indicators of blood loss, including deterioration in vital signs and reductions in hemoglobin are typically delayed in their presentation. Much of this is due to the remarkable compensation of human physiology. It is only when compensation begins to fail that we see the typical changes in vital signs we associate with hemorrhage. These changes are well described by the American College of Surgeons Advanced Trauma Life Support (ATLS) classification of blood loss. Using this system, heart rate only starts to increase to $>$100 beats per minute when blood loss reaches 15-30\% of total blood volume or 750-1500 mL, and MAP only begins to decline when blood loss reaches 30-40\% of total blood volume, or 1500-2000mL \citep{ACS_ATLS_10th}. With the failure of physiological compensation, hypoperfusion, end-organ injury and death rapidly ensue. Early detection and treatment of bleeding events has the potential to interrupt the downward spiral of hemorrhagic shock, prevent organ failure, reduce death, and limit total transfusion requirements. These implications are meaningful to patients and important from the standpoint of resource utilization. To make strides in this area, it is clear that we need tools and early warning systems that outperform clinicians monitoring basic vital signs and changes in hemoglobin. We hope that leveraging novel statistical frameworks will facilitate better identification and earlier intervention for this challenging problem.

The current manuscript presents a retrospective study on existing patient encounters; the immediate next step for future work is to produce an interpretable \texttt{R} Shiny Application with a user-interface tailored to clinical use in a real-time ICU setting. The goal is to develop an RSM that will quantify uncertainty about various medical conditions that are inherently difficult to identify in real-time, more than just internal bleeding. Additionally, the mean structure of our conditional response is clinically advisable, however, our approach requires assuming that the shock phenotype for an individual is the same over time. In other words, if a patient suffers from two distinct hemorrhage events during their ICU encounter, the current model assumes that the expected change in vitals per time instance will be the same for both bleeding events; physiologically, this need not be the case. A more flexible characterization of the latent states is worth exploring next.

\if1\blind
{
\section{Funding}
Research reported in this publication was supported by the National Heart, Lung, and Blood Institute of the National Institutes of Health under Award Number R56HL155373.
}\fi

{
\spacingset{1}
\footnotesize
\bibliographystyle{agsm}
\bibliography{References}
}%

%% file: supp.tex
\setcounter{page}{1}

\spacingset{1.25}

\section{More Details on Electronic Health Record Data}\label{chap3:app:data}
\subsection{Data Cleaning}
There are 33 distinct response measurements for each of the 33,924 patients, and the length of each patient visit varies between 15 minutes and 50 days (the average length of stay is approximately 35 hours). Despite the large electronic health record (EHR) database available, not all of the information is necessary to analyze. To start, all 33 distinct response outcomes are not necessary to incorporate into the model because not all are relevant for the detection of hemorrhage. Based on clinical expertise, the relevant EHR measurements for hemorrhage detection are: hemoglobin concentration, heart rate, mean arterial pressure (MAP), lactate levels, red blood cell (RBC) transfusion order and administration times, and patient medications. In particular, hemoglobin, heart rate, MAP, and lactate serve as a four-dimensional longitudinal response, while the RBC transfusions and medication information serve as covariates. Given the response and covariate information, we then clean the training data based on the following criteria: (1) the patient did not die during their encounter; (2) the patient encounter lasts at least six hours; (3) if the patient encounter lasts more than 48 hours, only the first 48 hours are used; (4) less than 10\% of heart rate and MAP measurements are missing; (5) patients do not have a pacemaker; (6) erroneous data points are removed (i.e., $\text{hemoglobin} \in [0,20]$, $\text{heart rate} \in [20,200]$, $\text{MAP} \in [25,150]$); (7) at least 80\% of the patient encounter is spent in intensive care unit (ICU) level-of-care. After adjusting for these criteria, the resulting dataset consists of 1,516 patient encounters. Due to the size of the data and model complexity, the compute time for training on all 1,516 subjects is computationally burdensome; hence, we use 500 subjects for training, and use five for testing. The reason for such a small test set is this allows us to have our clinical collaborators review these patient encounters closely and provide an in-depth commentary on the performance of our model with regards to uncovering their latent physiological changes, despite the data having no physiological-state labels.

\subsection{Medications}
The medication information most relevant for our use are: medication type, medication dose/strength, administration time, and administration type. With the help of clinical expertise, the medications can be separated based on the type of effect they are expected to have on heart rate and/or MAP. Similarly, we can separate medications based on if the administration type is \textit{continuous} (e.g., intravenous) or \textit{discrete} (e.g., oral pill). In total, there exists 44 distinct types of medications that are administered among all of the patients in our sample. Each medication can be separated based on how it is administered and how it affects heart rate and MAP, respectively. Hence, we have the following breakdown: 16 medications affect heart rate and are continuously administered, 18 medications affect heart rate and are discretely administered, 22 medications affect MAP and are continuously administered, and 28 medications affect MAP and are discretely administered. Thus, the 44 distinct medications leads to 84 medication effects to be learned in the model. Lastly, note that in the process of cleaning the medication data, any inconsistencies in a patient's medical record results in that patient being excluded from the training data. Section \ref{chap3:sec:method} in the manuscript details exactly how these medications are structured into the model. 

\section{Bayesian Computation Derivations} \label{chap3:app:bayes}
\noindent \fbox{\bf Full Conditional Distribution for $\Vec{(\balpha_*^{(i)})}$}
{\footnotesize
\begin{align*}
    \pi(\Vec{(\balpha_*^{(i)})} \mid \text{rest}) &\propto \pi\qty(\Vec{(\balpha_*^{(i)})} \mid \Vec{(\Tilde{\balpha}_*)}, \bUpsilon_\alpha)\\
    &\hspace{0.5in} \times \prod_{k=2}^{n_{i}}f\qty(\by^{(i)}_k \mid \by^{(i)}_{k-1}, \{\bb_j^{(i)} = s_{i,j}\}_{j=1}^k, \balpha_*^{(i)}, \bgamma^{(i)}, \bomega, \bbeta, \bA_{s_{i,k}}, \bR)\\
    &\propto \text{N}_{16}\qty(\Vec(\balpha_*^{(i)}) \mid \Vec(\Tilde{\balpha}_*),\; \bUpsilon_\alpha) \times \prod_{k=2}^{n_{i}} \text{N}_4\qty(\by^{(i)}_k \mid \bnu_{k}^{(i)} + \bA_{s_{i,k}}(\by^{(i)}_{k-1} - \bnu_{k-1}^{(i)}), \bR).
\end{align*}
}%
The product of the probability density functions above leads to the following distribution:
\begin{align*}
    \Vec{(\balpha_*^{(i)})} \mid \text{rest} &\sim \text{N}_{16} \qty(\vb*{W}_i \vb*{V}_i, \vb*{W}_i)
\end{align*}
where
{\footnotesize
\begin{align*}
    \vb*{W}_i &= \qty{\bUpsilon_\alpha^{-1} + \sum_{k=2}^{n_i} \qty(\bD_{\alpha, k}^{(i)} - \bA_{s_{i,k}} \bD_{\alpha, k-1}^{(i)})^{\top} \bR^{-1}\qty(\bD_{\alpha, k}^{(i)} - \bA_{s_{i,k}} \bD_{\alpha, k-1}^{(i)})}^{-1}\\
    \vb*{V}_i &= \bUpsilon_\alpha^{-1} \Vec(\Tilde{\balpha}_*) + \sum_{k=2}^{n_i} \qty(\bD_{\alpha, k}^{(i)} - \bA_{s_{i,k}} \bD_{\alpha, k-1}^{(i)})^{\top} \bR^{-1}\Big[\by^{(i)}_k - \bA_{s_{i,k}} \by^{(i)}_{k-1} - (\bD_{\omega, k}^{(i)} - \bA_{s_{i,k}} \bD_{\omega, k-1}^{(i)})\bomega\\
    &\hspace{3in} - (\bX_k^{(i)} - \bA_{s_{i,k}} \bX_{k-1}^{(i)})\bbeta - (\vb{I} - \bA_{s_{i,k}}) \bgamma^{(i)}\Big]
\end{align*}
}%

\noindent \fbox{\bf Full Conditional Distribution for $\bgamma^{(i)}$}
{\footnotesize
\begin{align*}
    \pi(\bgamma^{(i)} \mid \text{rest}) &\propto \pi\qty(\bgamma^{(i)} \mid \by^{(i)}_1, \bomega, \bbeta, \vb*{G}) \cdot f\qty(\by^{(i)}_1 \mid \bb_1^{(i)}=s_{i,1}, \balpha_*^{(i)}, \bgamma^{(i)}, \bomega,\bbeta, \bA_{s_{i,1}}, \bR)\\
    &\hspace{0.5in} \times \prod_{k=2}^{n_{i}}f\qty(\by^{(i)}_k \mid \by^{(i)}_{k-1}, \{\bb_j^{(i)} = s_{i,j}\}_{j=1}^k, \balpha_*^{(i)}, \bgamma^{(i)}, \bomega, \bbeta, \bA_{s_{i,k}}, \bR)\\
    &\propto \text{N}_4(\bgamma^{(i)} \mid \by^{(i)}_1 - \bD_{\omega, 1}^{(i)} \cdot \bomega - \bX_1^{(i)}\cdot \bbeta,\; \vb*{G}) \cdot \text{N}_4(\by^{(i)}_1 \mid \bgamma^{(i)} + \bD_{\omega, 1}^{(i)} \cdot \bomega + \bX_1^{(i)}\cdot \bbeta,\; \bGamma_{s_{i,1}})\\
    &\hspace{0.5in} \times \prod_{k=2}^{n_{i}} \text{N}_4\qty(\by^{(i)}_k \mid \bnu_{k}^{(i)} + \bA_{s_{i,k}}\cdot(\by^{(i)}_{k-1} - \bnu_{k-1}^{(i)}), \bR).
\end{align*}
}%
The product of the probability density functions above leads to the following distribution:
\begin{align*}
    \bgamma^{(i)} \mid \text{rest} &\sim \text{N}_{4} \qty(\vb*{W}_i \vb*{V}_i, \vb*{W}_i)
\end{align*}
where
{\footnotesize
\begin{align*}
    \vb*{W}_i &= \qty{ \vb*{G}^{-1} + \bGamma^{-1}_{s_{i,1}} + \sum_{k=2}^{n_i} \qty(\vb{I} - \bA_{s_{i,k}})^{\top} \bR^{-1}\qty(\vb{I} - \bA_{s_{i,k}})}^{-1}\\
    \vb*{V}_i &= \qty(\vb*{G}^{-1} + \bGamma^{-1}_{s_{i,1}})\qty(\by^{(i)}_1 - \bD_{\omega, 1}^{(i)} \cdot \bomega - \bX_1^{(i)}\cdot \bbeta) + \\
    &\hspace{1in} \sum_{k=2}^{n_i} \qty(\vb{I} - \bA_{s_{i,k}})^{\top} \bR^{-1}\Big[\by^{(i)}_k - \bA_{s_{i,k}} \by^{(i)}_{k-1} - \qty(\bD_{\alpha, k}^{(i)} - \bA_{s_{i,k}} \bD_{\alpha, k-1}^{(i)})\Vec(\balpha_*^{(i)})\\
    &\hspace{3in}- (\bD_{\omega, k}^{(i)} - \bA_{s_{i,k}} \bD_{\omega, k-1}^{(i)})\bomega - (\bX_k^{(i)} - \bA_{s_{i,k}} \bX_{k-1}^{(i)})\bbeta\Big]
\end{align*}
}%

\noindent \fbox{\bf Full Conditional Distribution for $\bomega$}
{\footnotesize
\begin{align*}
    \pi(\bomega \mid \text{rest}) &\propto \Bigg\{\prod_{i=1}^N \pi\qty(\bgamma^{(i)} \mid \by^{(i)}_1, \bomega, \bbeta, \vb*{G}) \cdot f\qty(\by^{(i)}_1 \mid \bb_1^{(i)}=s_{i,1}, \balpha_*^{(i)}, \bgamma^{(i)}, \bomega,\bbeta, \bA_{s_{i,1}}, \bR)\\
    &\hspace{0.15in} \times \prod_{k=2}^{n_{i}}f\qty(\by^{(i)}_k \mid \by^{(i)}_{k-1}, \{\bb_j^{(i)} = s_{i,j}\}_{j=1}^k, \balpha_*^{(i)}, \bgamma^{(i)}, \bomega, \bbeta, \bA_{s_{i,k}}, \bR)\Bigg\} \cdot \pi\qty(\bomega \mid \bomega_0, \bSigma_\omega)\\
    &\propto \Bigg\{\prod_{i=1}^N \text{N}_4(\bgamma^{(i)} \mid \by^{(i)}_1 - \bD_{\omega, 1}^{(i)} \cdot \bomega - \bX_1^{(i)}\cdot \bbeta,\; \vb*{G}) \cdot \text{N}_4(\by^{(i)}_1 \mid \bgamma^{(i)} + \bD_{\omega, 1}^{(i)} \cdot \bomega + \bX_1^{(i)}\cdot \bbeta,\; \bGamma_{s_{i,1}}) \\
    &\hspace{0.5in} \times \prod_{k=2}^{n_{i}} \text{N}_4\qty(\by^{(i)}_k \mid \bnu_{k}^{(i)} + \bA_{s_{i,k}}\cdot(\by^{(i)}_{k-1} - \bnu_{k-1}^{(i)}), \bR)\Bigg\} \cdot \text{N}_{84}\qty(\bomega \mid \bomega_0, \bSigma_\omega).
\end{align*}
}%
The product of the probability density functions above leads to the following distribution:
\begin{align*}
    \bomega \mid \text{rest} &\sim \text{N}_{84} \qty(\vb*{W} \vb*{V}, \vb*{W})
\end{align*}
where
{\footnotesize
\begin{align*}
    \vb*{W} &= \qty{\bSigma_\omega^{-1} + \sum_{i=1}^N{\bD_{\omega, 1}^{(i)}}^{\top} \qty(\vb*{G}^{-1} + \bGamma_{s_{i,1}}^{-1}) \bD_{\omega, 1}^{(i)} + \sum_{k=2}^{n_i} \qty(\bD_{\omega, k}^{(i)} - \bA_{s_{i,k}} \bD_{\omega, k-1}^{(i)})^{\top} \bR^{-1}\qty(\bD_{\omega, k}^{(i)} - \bA_{s_{i,k}} \bD_{\omega, k-1}^{(i)})}^{-1}\\
    \vb*{V} &= \bSigma_\omega^{-1}\bomega_0 + \sum_{i=1}^{N} {\bD_{\omega, 1}^{(i)}}^{\top} \qty(\vb*{G}^{-1} + \bGamma_{s_{i,1}}^{-1}) \qty(\by^{(i)}_1 - \bgamma^{(i)} - \bX_1^{(i)} \cdot \bbeta)\\
    &\quad + \sum_{k=2}^{n_i} \qty(\bD_{\omega, k}^{(i)} - \bA_{s_{i,k}} \bD_{\omega, k-1}^{(i)})^{\top} \bR^{-1}\Big[\by^{(i)}_k - \bA_{s_{i,k}} \by^{(i)}_{k-1} - (\bD_{\alpha, k}^{(i)} - \bA_{s_{i,k}} \bD_{\alpha, k-1}^{(i)})\Vec(\balpha_*^{(i)})\\
    &\hspace{2in} \quad - (\bX_k^{(i)} - \bA_{s_{i,k}} \bX_{k-1}^{(i)})\bbeta - (\vb{I} - \bA_{s_{i,k}}) \bgamma^{(i)}\Big]
\end{align*}
}%

\noindent \fbox{\bf Full Conditional Distribution for $\bbeta$}
{\footnotesize
\begin{align*}
    \pi(\bbeta \mid \text{rest}) &\propto \Bigg\{\prod_{i=1}^N \pi\qty(\bgamma^{(i)} \mid \by^{(i)}_1, \bomega, \bbeta, \vb*{G}) \cdot f\qty(\by^{(i)}_1 \mid \bb_1^{(i)}=s_{i,1}, \balpha_*^{(i)}, \bgamma^{(i)}, \bomega,\bbeta, \bA_{s_{i,1}}, \bR)\\
    &\hspace{0.15in} \times \prod_{k=2}^{n_{i}}f\qty(\by^{(i)}_k \mid \by^{(i)}_{k-1}, \{\bb_j^{(i)} = s_{i,j}\}_{j=1}^k, \balpha_*^{(i)}, \bgamma^{(i)}, \bomega, \bbeta, \bA_{s_{i,k}}, \bR)\Bigg\} \cdot \pi\qty(\bbeta \mid \bbeta_0, \bSigma_\beta)\\
    &\propto \Bigg\{\prod_{i=1}^N \text{N}_4(\bgamma^{(i)} \mid \by^{(i)}_1 - \bD_{\omega, 1}^{(i)} \cdot \bomega - \bX_1^{(i)}\cdot \bbeta,\; \vb*{G}) \cdot \text{N}_4(\by^{(i)}_1 \mid \bgamma^{(i)} + \bD_{\omega, 1}^{(i)} \cdot \bomega + \bX_1^{(i)}\cdot \bbeta,\; \bGamma_{s_{i,1}}) \\
    &\hspace{0.5in} \times \prod_{k=2}^{n_{i}} \text{N}_4\qty(\by^{(i)}_k \mid \bnu_{k}^{(i)} + \bA_{s_{i,k}}\cdot(\by^{(i)}_{k-1} - \bnu_{k-1}^{(i)}), \bR)\Bigg\} \cdot \text{N}_{4}\qty(\bbeta \mid \bbeta_0, \bSigma_\beta).
\end{align*}
}%
The product of the probability density functions above leads to the following distribution:
\begin{align*}
    \bbeta \mid \text{rest} &\sim \text{N}_{4} \qty(\vb*{W} \vb*{V}, \vb*{W})
\end{align*}
where
{\footnotesize
\begin{align*}
    \vb*{W} &= \qty{\bSigma_\beta^{-1} + \sum_{i=1}^N {\bX^{(i)}_1}^{\top} \qty(\vb*{G}^{-1} + \bGamma_{s_{i,1}}^{-1})\bX^{(i)}_1 + \sum_{k=2}^{n_i} \qty(\bX^{(i)}_k - \bA_{s_{i,k}}\bX^{(i)}_{k-1})^{\top}\bR^{-1}\qty(\bX^{(i)}_k - \bA_{s_{i,k}}\bX^{(i)}_{k-1})}^{-1}\\
    \vb*{V} &= \bSigma_\beta^{-1}\bbeta_0 + \sum_{i=1}^N {\bX^{(i)}_1}^{\top} \qty(\vb*{G}^{-1} + \bGamma_{s_{i,1}}^{-1})\qty(\by^{(i)}_1 - \bgamma^{(i)} - \bD_{\omega, 1}^{(i)}\bomega)\\
    &\quad  + \sum_{k=2}^{n_i} (\bX_k^{(i)} - \bA_{s_{i,k}} \bX_{k-1}^{(i)})^{\top} \bR^{-1}\Big[\by^{(i)}_k - \bA_{s_{i,k}} \by^{(i)}_{k-1} - (\bD_{\alpha, k}^{(i)} - \bA_{s_{i,k}} \bD_{\alpha, k-1}^{(i)})\Vec(\balpha_*^{(i)})\\
    &\hspace{2in} \quad - (\bD_{\omega, k}^{(i)} - \bA_{s_{i,k}} \bD_{\omega, k-1}^{(i)})\bomega - (\vb{I} - \bA_{s_{i,k}}) \bgamma^{(i)}\Big]
\end{align*}
}%

\noindent \fbox{\bf Full Conditional Distribution for $\Vec(\Tilde{\balpha}_*)$}
\begin{align*}
    \pi(\Vec{(\Tilde{\balpha}_*)} \mid \text{rest}) &\propto \text{N}_{16}(\Vec(\Tilde{\balpha}_*) \mid \Vec(\Tilde{\balpha}_0), \bSigma_\alpha) \cdot \prod_{i=1}^N \text{N}_{16}(\Vec{(\balpha_*^{(i)})} \mid \Vec{(\Tilde{\balpha}_*)}, \bUpsilon_\alpha)
\end{align*}
The product of the probability density functions above leads to the following distribution:
\begin{align*}
    \Vec{(\Tilde{\balpha}_*)} \mid \text{rest} &\sim \text{N}_{20} \qty(\vb*{W} \vb*{V}, \vb*{W})
\end{align*}
where
\begin{align*}
    \vb*{W} &= \qty(\bSigma_\alpha^{-1} + N\cdot \bUpsilon_{\alpha}^{-1})^{-1}\\
    \vb*{V} &= \bSigma_\alpha^{-1} \Vec(\Tilde{\balpha}_0) + \bUpsilon_\alpha^{-1}\cdot \sum_{i=1}^N \Vec(\balpha_*^{(i)})
\end{align*}

\noindent \fbox{\bf Full Conditional Distribution for $\bUpsilon_\alpha$}
\begin{align*}
    \pi(\bUpsilon_\alpha \mid \text{rest}) &\propto \qty{\prod_{i=1}^N \text{N}_{16}\qty(\Vec{(\balpha_*^{(i)})} \mid \Vec{(\Tilde{\balpha}_*)}, \bUpsilon_\alpha)} \cdot \text{InvWish}\qty(\bUpsilon_\alpha \mid \bPsi_\alpha, \nu_\alpha)
\end{align*}
The product of the probability density functions above leads to the following distribution:
\begin{align*}
    \bUpsilon_\alpha \mid \text{rest} \sim \text{InvWish}\qty(\bPsi_\alpha + \sum_{i=1}^N\qty[\Vec(\balpha_*^{(i)}) - \Vec(\Tilde{\balpha}_*)]\qty[\Vec(\balpha_*^{(i)}) - \Vec(\Tilde{\balpha}_*)]^{\top}, \nu_\alpha + N)
\end{align*}

\noindent \fbox{\bf Full Conditional Distribution for $\vb*{G}$}
\begin{align*}
    \pi(\vb*{G} \mid \text{rest}) &\propto \qty{\prod_{i=1}^N \text{N}_4(\bgamma^{(i)} \mid \by^{(i)}_1 - \bD_{\omega, 1}^{(i)} \cdot \bomega - \bX_1^{(i)}\cdot \bbeta,\; \vb*{G})} \cdot \text{InvWish}\qty(\vb*{G} \mid \bPsi_G, \nu_G)
\end{align*}
The product of the probability density functions above leads to the following distribution:
{\footnotesize
\begin{align*}
    \vb*{G} \mid \text{rest} \sim \text{InvWish}\qty(\bPsi_G + \sum_{i=1}^N \qty[\bgamma^{(i)} - \by^{(i)}_1 + \bD_{\omega, 1}^{(i)} \cdot \bomega + \bX_1^{(i)}\cdot \bbeta]\qty[\bgamma^{(i)} - \by^{(i)}_1 + \bD_{\omega, 1}^{(i)} \cdot \bomega + \bX_1^{(i)}\cdot \bbeta]^{\top}, \nu_G + N)
\end{align*}
}%

\noindent \fbox{\bf ``Approximate-Gibbs'' Metropolis-Hastings Update for $\bR$}\\
Because of the autoregressive nature of our model, the conditional distribution of the initial time point has error covariance $\bGamma_{s_{i,1}}$ instead of $\bR$ [which the remaining time points use for the error covariance]. As a result, deriving a Gibbs update for $\bR$ is infeasible; however, because $\bGamma_{s_{i,1}}$ is a function of $\bR$, we can derive an ``approximate Gibbs'' update for $\bR$ which is in theory a Metropolis-Hastings update, but the proposal distribution emulates that of a full conditional distribution for $\bR$. In particular, the full conditional distribution for $\bR$ is proportional to the following
\begin{align*}
    \pi(\bR \mid \text{rest}) &\propto \pi(\bR \mid \vb*{\Psi}_R, \nu_R) \cdot \prod_{i=1}^N f\qty(\by^{(i)}_1 \mid \bb_1^{(i)}=s_{i,1}, \balpha_*^{(i)}, \bgamma^{(i)}, \bomega,\bbeta, \bA_{s_{i,1}}, \bR)\\
    &\hspace{0.5in} \times \prod_{k=2}^{n_{i}}f\qty(\by^{(i)}_k \mid \by^{(i)}_{k-1}, \{\bb_j^{(i)} = s_{i,j}\}_{j=1}^k, \balpha_*^{(i)}, \bgamma^{(i)}, \bomega, \bbeta, \bA_{s_{i,k}}, \bR)\\
    &\propto \text{InvWish}\qty(\bR \mid \vb*{\Psi}_R, \nu_R) \cdot\prod_{i=1}^N \text{N}_4(\by^{(i)}_1 \mid \bgamma^{(i)} + \bD_{\omega, 1}^{(i)} \cdot \bomega + \bX_1^{(i)}\cdot \bbeta,\; \bGamma_{s_{i,1}}) \\
    &\hspace{0.5in} \times \prod_{k=2}^{n_{i}} \text{N}_4\qty(\by^{(i)}_k \mid \bnu_{k}^{(i)} + \bA_{s_{i,k}}\cdot(\by^{(i)}_{k-1} - \bnu_{k-1}^{(i)}), \bR).
\end{align*}
Now, define $\vb*{B}_i$ to be a matrix such that $\vb*{B}_i\bGamma_{s_{i,1}}\vb*{B}_i^{\top} = \bR$. Then, $\vb*{B}_i = \bR^{1/2} \bGamma_{s_{i,1}}^{-1/2}$. As a result, we know the following:
\begin{align*}
    \vb*{B}_i\cdot \by^{(i)}_1 &\sim \text{N}_4\qty(\vb*{B}_i\cdot \qty[\bgamma^{(i)} + \bD_{\omega, 1}^{(i)} \cdot \bomega + \bX_1^{(i)}\cdot \bbeta], \bR)
\end{align*}
Therefore, we get the following
\begin{align*}
    \pi(\bR \mid \text{rest}) &\approx \text{InvWish}\qty(\bR \mid \vb*{\Psi}_R, \nu_R) \cdot\prod_{i=1}^N\text{N}_4\qty(\vb*{B}_i\cdot \by^{(i)}_1 \mid \vb*{B}_i\cdot \qty[\bgamma^{(i)} + \bD_{\omega, 1}^{(i)} \cdot \bomega + \bX_1^{(i)}\cdot \bbeta], \bR)\\
    &\hspace{0.5in} \times \prod_{k=2}^{n_{i}} \text{N}_4\qty(\by^{(i)}_k \mid \bnu_{k}^{(i)} + \bA_{s_{i,k}}\cdot(\by^{(i)}_{k-1} - \bnu_{k-1}^{(i)}), \bR).
\end{align*}
The approximate full conditional distribution for $\bR$ above simplifies to an Inverse-Wishart distribution dependent on the value of $\vb*{B}_i = \bR^{1/2} \bGamma_{s_{i,1}}^{-1/2}$. Therefore, given the current value of $\vb*{B}_i$ (i.e., $\bR$), we can write the proposal distribution for some new $\bR^*$ as
\begin{align*}
    q(\bR^* \mid \bR, \text{rest}) \sim \text{InvWish}(\bPsi_q, \nu_q)
\end{align*}
where 
\begin{align*}
    \bPsi_q &= \bPsi_R + \sum_{i=1}^N (\bR^{1/2} \bGamma_{s_{i,1}}^{-1/2})(\by^{(i)}_1 - \bnu_1^{(i)})(\by^{(i)}_1 - \bnu_1^{(i)})^{\top} (\bR^{1/2} \bGamma_{s_{i,1}}^{-1/2})^{\top}\\
    &\hspace{0.25in}+ \sum_{k=2}^{n_i} \qty[\by^{(i)}_k - \bnu_{k}^{(i)} - \bA_{s_{i,k}}\cdot(\by^{(i)}_{k-1} - \bnu_{k-1}^{(i)})]\qty[\by^{(i)}_k - \bnu_{k}^{(i)} - \bA_{s_{i,k}}\cdot(\by^{(i)}_{k-1} - \bnu_{k-1}^{(i)})]^{\top}\\
    \nu_q &= \nu_R + \sum_{i=1}^N n_i.
\end{align*}
where $\bnu_1^{(i)} = \bgamma^{(i)} + \bD_{\omega, 1}^{(i)} \cdot \bomega + \bX_1^{(i)}\cdot \bbeta$.

\section{Exact Prior Specifications}\label{chap3:app:priorSpec}
\noindent \fbox{\bf $\bA_1, \hdots, \bA_5$}
$$\vb{A}_j = \mqty(a_{1,j}&0&0&0\\ 0&a_{2,j}&0&0\\ 0&0&a_{3,j}&0\\ 0&0&0& a_{4,j}),$$
and 
$$(\text{logit}(a_{1,j}),\text{logit}(a_{2,j}),\text{logit}(a_{3,j}),\text{logit}(a_{4,j})) \overset{iid}{\sim} \text{N}_4(\vb*{0}, \vb*{I}),$$
for $j \in \{1,2,3,4,5\}$.

\noindent \fbox{\bf $\bR$}
$$\vb*{R} \sim \text{InvWish}(\bPsi_R, \nu_R)$$
where
$$\bPsi_R = \nu_R \cdot \mqty(\frac{1}{2} & 0 & 0 & 0\\ 0 & \frac{3}{2} & 0 & 0\\ 0 & 0 & \frac{3}{2} & 0\\ 0 & 0 & 0 & \frac{1}{2})$$
and $\nu_R = 2 \cdot \sum_{i=1}^N n_i$.

\noindent \fbox{\bf $\bzeta$}
$$\bzeta \sim \text{N}_{24}(\bmu_\zeta, \bSigma_\zeta)$$
where $\bmu_\zeta =$ (-7.2405, 2.5, -6.2152,   1, -2.6473,  -1, -6.1475,  -1, -9.4459,  -1, -7.2404, 2.5, -7.2151,   1, -7.1778, 2.5, -5.2151,   0, -9.4459,  -1, -7.2404, 2.5, -5.2151,   0), and $\bSigma_\zeta = \vb*{I}_{24}$.

\noindent \fbox{\bf $\bpi$}
$$\text{logit}(\bpi) \sim \text{N}_4(\vb*{0}, 100 \cdot \vb*{I})$$

\noindent \fbox{\bf $\vb*{\omega}$}
$$\bomega \sim \text{N}_{84}(\bomega_0,\; \bSigma_\omega)$$
where 
\begin{align*}
    \bomega_0 = \frac{3}{2}\cdot (&-1, 1, 1,-1,-1, 1, 1,-1, 1, 1,-1,-1, 1, -1, 1, 1,-1,-1,-1,-1, 1,-1, 1,-1, 1,\\
                 &-1,-1,-1,-1,-1, 1, 1,-1,-1,-1,-1,-1, 1, 1, 1,-1, 1,-1,-1,-1, 1,-1, 1,\\
                 &-1,1,-1,-1,-1, 1, 1,-1,-1,-1,-1,-1,-1,-1,-1,-1,-1, 1, 1, 1,-1,-1,\\
                 &-1, 1,-1,1,-1,-1,-1,-1, 1,-1,-1,-1,-1,-1),
\end{align*}
$\bSigma_\omega = \frac{1}{16} \cdot \vb*{I}_{84}$, and $[\bSigma_\omega]_{6,6} = [\bSigma_\omega]_{42,42} = [\bSigma_\omega]_{49,49} = \frac{1}{625}$.

\noindent \fbox{\bf $\vb*{\beta}$}
$$\bbeta \sim \text{N}_{4}(\bbeta_0,\; \bSigma_\beta)$$
$\bbeta_0 = \qty(\frac{1}{4}, -2, 2, -\frac{1}{4})$ and 
$$\bSigma_\beta = \mqty(1 & 0 & 0 & 0\\ 0 & 16 & 0 & 0\\ 0 & 0 & 16 & 0\\ 0 & 0 & 0 & 16)$$

\noindent \fbox{\bf $\bUpsilon_\alpha$}
$$\bUpsilon_\alpha \sim \text{InvWish}(\vb*{\Psi}_\alpha,\; \nu_\alpha)$$
where $$\bPsi_\alpha = (\nu_\alpha - 16 - 1)\cdot \text{diag}\qty(\frac{1}{4}, \frac{1}{4},  4,  4,\frac{9}{4}, \frac{9}{4}, 25, 25,\frac{9}{4}, \frac{9}{4}, 25, 25,\frac{1}{4}, \frac{1}{4},  4,  4)$$
and $\nu_\alpha = 80$.

\noindent \fbox{\bf $\Vec(\Tilde{\balpha}_*)$}
$$\Vec(\Tilde{\balpha}_*) \sim \text{N}_{16}(\Vec(\Tilde{\balpha}_0),\; \bSigma_\alpha)$$
where 
\begin{align*}
    \Vec(\Tilde{\balpha}_0) &= (-1,  1, 0, 0, 7, -7, 0, 0, -7,  7, 0, 0, 1, -1, 0, 0),\\
    \bSigma_\alpha &= \text{diag}\qty(\frac{1}{4},\frac{1}{4},4,4,4,4,64,64,4,4,64,64,\frac{1}{4},\frac{1}{4},4,4)
\end{align*}

\noindent \fbox{\bf $\vb*{G}$}
$$\vb*{G} \sim \text{InvWish}(\vb*{\Psi}_G,\; \nu_G)$$
where $\vb*{\Psi}_G = (\nu_G - 4 - 1)\cdot  \text{diag}(8, 32, 32, 8)$ and $\nu_G = 8$.

\section{Data Imputation}
Since we have missingness in the data, data imputation for the response vector is necessary. The following are the distributions from which we sample the missing observations from.

\noindent\fbox{$k=1$}
\begin{align*}
    \pi(\by^{(i)}_1 \mid \text{rest}) &\propto \pi\qty(\bgamma^{(i)} \mid \by^{(i)}_1, \bomega, \bbeta, \vb*{G}) \cdot f\qty(\by^{(i)}_1 \mid \bb_1^{(i)}=s_{i,1}, \balpha_*^{(i)}, \bgamma^{(i)}, \bomega,\bbeta, \bA_{s_{i,1}}, \bR)\\
    &\hspace{0.5in} \times f\qty(\by^{(i)}_2 \mid \by^{(i)}_{1}, \{\bb_j^{(i)} = s_{i,j}\}_{j=1}^2, \balpha_*^{(i)}, \bgamma^{(i)}, \bomega, \bbeta, \bA_{s_{i,2}}, \bR).
\end{align*}
The product of the probability density functions above leads to the following distribution
\begin{align*}
    \by^{(i)}_1 \mid \text{rest} &\sim \text{N}_{4} \qty(\vb*{W}_{i,1} \vb*{V}_{i,1}, \vb*{W}_{i,1})
\end{align*}
where
{\footnotesize
\begin{align*}
    \vb*{W}_{i,1} &= \qty(\vb*{G}^{-1} + \bGamma_{s_{i,1}}^{-1} + \bA_{s_{i,2}}^{\top}\bR^{-1}\bA_{s_{i,2}})^{-1}\\
    \vb*{V}_{i,1} &= \qty(\vb*{G}^{-1} + \bGamma_{s_{i,1}}^{-1})\qty(\bgamma^{(i)} + \bD_{\omega, 1}^{(i)} \cdot \bomega + \bX_1^{(i)}\cdot \bbeta)\\
    &\hspace{0.1in} + \bA_{s_{i,2}}^{\top} \bR^{-1}\qty(\by^{(i)}_2 - \bD_{\alpha, 2}^{(i)} \Vec(\balpha_*^{(i)}) - (\bD_{\omega, 2}^{(i)} - \bA_{s_{i,2}} \bD_{\omega, 1}^{(i)})\bomega - (\bX_2^{(i)} - \bA_{s_{i,2}} \bX_{1}^{(i)})\bbeta - (\vb{I} - \bA_{s_{i,2}}) \bgamma^{(i)}).
\end{align*}
}%

\noindent\fbox{$k \in \{2,3,\hdots, n_i - 1\}$}
\begin{align*}
    \pi(\by^{(i)}_k \mid \text{rest}) &\propto f\qty(\by^{(i)}_k \mid \by^{(i)}_{k-1}, \{\bb_j^{(i)} = s_{i,j}\}_{j=1}^k, \balpha_*^{(i)}, \bgamma^{(i)}, \bomega, \bbeta, \bA_{s_{i,k}}, \bR)\\
    &\hspace{0.5in} \times f\qty(\by^{(i)}_{k+1} \mid \by^{(i)}_{k}, \{\bb_j^{(i)} = s_{i,j}\}_{j=1}^{k+1}, \balpha_*^{(i)}, \bgamma^{(i)}, \bomega, \bbeta, \bA_{s_{i,k+1}}, \bR).
\end{align*}
The product of the probability density functions above leads to the following distribution
\begin{align*}
    \by^{(i)}_k \mid \text{rest} &\sim \text{N}_{4} \qty(\vb*{W}_{i,k} \vb*{V}_{i,k}, \vb*{W}_{i,k})
\end{align*}
where
{\footnotesize
\begin{align*}
    \vb*{W}_{i,k} &= \qty(\bR^{-1} + \bA_{s_{i,k+1}}^{\top}\bR^{-1}\bA_{s_{i,k+1}})^{-1}\\
    \vb*{V}_{i,k} &= \bR^{-1}\cdot \Big[\bA_{s_{i,k}} \by^{(i)}_{k-1} + (\bD_{\alpha, k}^{(i)} - \bA_{s_{i,k}} \bD_{\alpha, k-1}^{(i)})\Vec(\balpha_*^{(i)}) + (\bD_{\omega, k}^{(i)} - \bA_{s_{i,k}} \bD_{\omega, k-1}^{(i)})\bomega\\
    &\hspace{0.2in}+ (\bX_k^{(i)} - \bA_{s_{i,k}} \bX_{k-1}^{(i)}) \bbeta + (\vb{I} - \bA_{s_{i,k}}) \bgamma^{(i)}\Big] \\
    &+ \bA_{s_{i,k+1}}^{\top}\bR^{-1} \cdot \Big[\by^{(i)}_{k+1} - (\bD_{\alpha, k+1}^{(i)} - \bA_{s_{i,k+1}} \bD_{\alpha, k}^{(i)})\Vec(\balpha_*^{(i)}) - (\bD_{\omega, k+1}^{(i)} - \bA_{s_{i,k+1}} \bD_{\omega, k}^{(i)})\bomega\\
    &\hspace{0.2in}- (\bX_{k+1}^{(i)} - \bA_{s_{i,k+1}} \bX_{k}^{(i)}) \bbeta - (\vb{I} - \bA_{s_{i,k+1}}) \bgamma^{(i)}\Big].
\end{align*}
}%

\noindent \fbox{$k=n_i$}
\begin{align*}
    \pi(\by^{(i)}_{n_i} \mid \text{rest}) &\propto f\qty(\by^{(i)}_{n_i} \mid \by^{(i)}_{n_i-1}, \{\bb_j^{(i)} = s_{i,j}\}_{j=1}^{n_i}, \balpha_*^{(i)}, \bgamma^{(i)}, \bomega, \bbeta, \bA_{s_{i,n_i}}, \bR).
\end{align*}
Thus, we trivially get the conditional response distribution for $\by_{n_i}^{(i)}$:
\begin{align*}
    \by^{(i)}_{n_i} \mid \text{rest} &\sim \text{N}_{4} \qty(\bnu^{(i)}_{n_i} + \bA(\by^{(i)}_{n_i -1} - \bnu^{(i)}_{n_i-1}), \bR)
\end{align*}

\pagebreak

\section{Our Metropolis-Hastings (MH) State-Sampler} \label{chap3:app:stateSampAlgs}
{
\spacingset{1.3}
\begin{algorithm}[!htb]
\footnotesize
\DontPrintSemicolon
    \KwInput{Current $\{\bb_j^{(i)} = s_{i,j}\}_{j=1}^{n_i}$ and model parameters}
    \KwData{$\vb*{Y}^{(i)}$ for a given $i \in \{1,2,\hdots, N\}$}
    
    $p = $ randomly select from $\{2, 3, \hdots, 50\}$ with equal probability\\
    \If{$p \geq n_i$} {
        Use Algorithm \ref{chap3:alg:statesamp2}
    } \ElseIf{$p=2$} {
        Use Algorithm \ref{chap3:alg:statesamp3}
    } \Else {
        $k = 1$ \\
        \While{$k \leq n_i - 2$} { 
            Initialize $s_{i,j}^* = s_{i,j}$, $\forall j \in \{1,2,\hdots, n_i\}$\\
            $k_{\text{max}} = \min\{k+p-1, n_i\}$\\
            \tcc{Propose state sequence $\{\bb_j^{(i)} = s_{i,j}^*\}_{j=k}^{k_{\text{max}}}$ for time points $k, k+1, \hdots, k_{\text{max}}$}
            \For{$t \in \{k, k+1, \hdots, k_{\text{max}}-2\}$} {
                \texttt{like\_vals}: vector with length equal to number of distinct possible states (i.e., size of state-space)\\
                \For{$m \in \{1,2,\hdots, 5\}$} {
                    \If{$t = 1$} {
                        \texttt{like\_vals}[$m$] = $\bpi_{m} \cdot f\qty(\by^{(i)}_1 \mid \bb_1^{(i)}=m, \balpha_*^{(i)}, \bgamma^{(i)}, \bomega,\bbeta, \bA_{m}, \bR)$
                    } \Else {
                        \texttt{like\_vals}[$m$] = $\vb{P}_{s_{i,t-1}^*,m} \cdot f\qty(\by^{(i)}_t \mid \by^{(i)}_{t-1}, \{\bb_j^{(i)} = s_{i,j}^*\}_{j=1}^{t-1}, \bb_t^{(i)} = m, \balpha_*^{(i)}, \bgamma^{(i)}, \bomega, \bbeta, \bA_{m}, \bR)$
                    }
                }
        
                $s_{i,t}^* = $ sample from $\{1,2,\hdots,5\}$ proportional to \texttt{like\_vals}
            }
            
            \If{$k_{\text{max}} < n_i$} {
                Randomly select $\{s_{i,k_{\text{max}}-1}^*, s_{i,k_{\text{max}}}^*\} \in \mathcal{B}^{(2)}_{s_{i,k_{\text{max}}-2}^*, s_{i,k_{\text{max}}+1}^*}$ with equal probability
            } \Else {
                Randomly select $\{s_{i,k_{\text{max}}-1}^*, s_{i,k_{\text{max}}}^*\} \in \mathcal{B}^{(2)}_{s_{i,k_{\text{max}}-2}^*, \boldsymbol{\cdot}}$ with equal probability
            }
            $a = $ MH acceptance ratio between the $\{\bb_j^{(i)} = s_{i,j}\}_{j=1}^{n_i}$ and $\{\bb_j^{(i)} = s^*_{i,j}\}_{j=1}^{n_i}$
        
            $u = $ sample from uniform$[0,1]$
            
            \If{$u < a$} {
                $s_{i,j} = s_{i,j}^*$, $\forall j \in \{1,2,\hdots, n_i\}$ \tcp*{proposal accepted}
            }
            $k = k + p -2$
        }
    
        \Return{$\{\bb_j^{(i)} = s_{i,j}\}_{j=1}^{n_i}$}
    }

\caption{MH state-sampling for subject $i\in\{1,2,\hdots, N\}$}
\label{chap3:alg:statesamp}
\end{algorithm}
}%

{
\spacingset{1.25}
\begin{algorithm}[!htb]
\footnotesize
\DontPrintSemicolon
    \KwInput{Current $\{\bb_j^{(i)} = s_{i,j}\}_{j=1}^{n_i}$ and model parameters}
    \KwData{$\vb*{Y}^{(i)}$ for a given $i \in \{1,2,\hdots, N\}$}
    \tcc{Proposing a full state sequence $\{\bb_j^{(i)} = s_{i,j}^*\}_{j=1}^{n_i}$}
    \For{$k \in \{1,2,\hdots, n_i\}$}    
    { 
    \texttt{like\_vals}: vector with length equal to number of distinct possible states (i.e., size of state-space)
        
        \For{$m \in \{1,2,\hdots, 5\}$}
        {
            \If{$k = 1$}
            {
                \texttt{like\_vals}[$m$] = $\bpi_{m} \cdot f\qty(\by^{(i)}_1 \mid \bb_1^{(i)}=m, \balpha_*^{(i)}, \bgamma^{(i)}, \bomega, \bbeta, \bA_m, \bR)$
            }
            \Else
            {
                \texttt{like\_vals}[$m$] = $\vb{P}_{s_{i,k-1}^*,m} \cdot f\qty(\by^{(i)}_k \mid \by^{(i)}_{k-1}, \{\bb_j^{(i)} = s_{i,j}^*\}_{j=1}^{k-1}, \bb_k^{(i)} = m, \balpha_*^{(i)}, \bgamma^{(i)}, \bomega, \bbeta, \bA_m, \bR)$
            }
        }

        $s_{i,k}^* = $ sample from $\{1,2,\hdots,5\}$ proportional to \texttt{like\_vals}
    }
    
    $a = $ MH acceptance ratio between the $\{\bb_j^{(i)} = s_{i,j}\}_{j=1}^{n_i}$ and $\{\bb_j^{(i)} = s^*_{i,j}\}_{j=1}^{n_i}$
    
    $u = $ sample from uniform$[0,1]$
    
    \If{$u < a$}
    {
        $s_{i,j} = s^*_{i,j}$, $\forall j \in \{1,2,\hdots, n_i\}$ \tcp*{proposal accepted}
    }

    \Return{$\{\bb_j^{(i)} = s_{i,j}\}_{j=1}^{n_i}$}

\caption{MH state-sampling for subject $i\in\{1,2,\hdots, N\}$ and $p \geq n_i$}
\label{chap3:alg:statesamp2}
\end{algorithm}
}%

{
\spacingset{1.25}
\begin{algorithm}[!htb]
\footnotesize
\DontPrintSemicolon
    \KwInput{Current $\{\bb_j^{(i)} = s_{i,j}\}_{j=1}^{n_i}$ and model parameters}
    \KwData{$\vb*{Y}^{(i)}$ for a given $i \in \{1,2,\hdots, N\}$}
    \For{$k \in \{1,2,\hdots, n_i\}$}    
    { 
        Initialize $s_{i,j}^* = s_{i,j}$, $\forall j \in \{1,2,\hdots, n_i\}$\\
        \tcc{Proposing $\{\bb_k^{(i)} = s_{i,k}^*, \bb_{k+1}^{(i)} = s_{i,k+1}^*\}$}
        \If{$k = 1$}
        {
            Randomly select $\{s_{i,1}^*, s_{i,2}^*\} \in \mathcal{B}^{(2)}_{\boldsymbol{\cdot}, s^*_{i,3}}$ with equal probability
        }
        \ElseIf{$k = n_i - 1$}
        {
            Randomly select $\{s_{i,n_i-1}^*, s_{i,n_i}^*\} \in \mathcal{B}^{(2)}_{s^*_{i,n_i-2}, \boldsymbol{\cdot}}$ with equal probability
        }
        \Else
        {
            Randomly select $\{s_{i,k+p-2}^*, s_{i,k+p-1}^*\} \in \mathcal{B}^{(2)}_{s_{i,k+p-3}^*, s_{i,k+p}^*}$ with equal probability
        }

        $a = $ MH acceptance ratio between the $\{\bb_j^{(i)} = s_{i,j}\}_{j=1}^{n_i}$ and $\{\bb_j^{(i)} = s^*_{i,j}\}_{j=1}^{n_i}$
    
        $u = $ sample from uniform$[0,1]$
        
        \If{$u < a$}
        {
            $s_{i,j} = s_{i,j}^*$, $\forall j \in \{1,2,\hdots, n_i\}$ \tcp*{proposal accepted}
        }
    }

    \Return{$\{\bb_j^{(i)} = s_{i,j}\}_{j=1}^{n_i}$}

\caption{MH state-sampling for subject $i\in\{1,2,\hdots, N\}$ and $p = 2$}
\label{chap3:alg:statesamp3}
\end{algorithm}
}%

\section{Additional Information on Algorithms \ref{chap3:alg:statesamp}, \ref{chap3:alg:statesamp2}, and \ref{chap3:alg:statesamp3}}\label{chap3:app:stateSampInfo}

Although Algorithms \ref{chap3:alg:statesamp}, \ref{chap3:alg:statesamp2}, and \ref{chap3:alg:statesamp3} provide pseudo-code for the logic of the state-sampling routine, the following provides further mathematical details for each of the algorithms. In particular, we define the proposal distributions for the cases of $p = 2$ and for $p \geq n_i$, for a given subject $i$. 

In the case of Algorithm \ref{chap3:alg:statesamp3} where $p=2$, the proposal distribution is precisely the same as \textit{Approach (A)} defined in Supplementary Materials Section \ref{chap3:app:altSamp}. In the case of Algorithm \ref{chap3:alg:statesamp2} where $p \geq n_i$, this more closely aligns with the proposal distribution provided in (\ref{chap3:eq:propDist}) in the manuscript. In particular, the proposal distribution for Algorithm \ref{chap3:alg:statesamp2} makes proposals over the entire state sequence for subject $i$, and is defined as
{
\footnotesize
\begin{align*}
    &q\qty(\{\vb*{b}_j^{(i)} = s_{i,j}\}_{j=1}^{n_i} \mid \vb*{Y}^{(i)}, \balpha_*^{(i)}, \bomega^{(i)}, \bgamma^{(i)}, \bbeta, \bA, \bR,\bzeta,\bpi)\\
    &\hspace{0.05in}:=
        \frac{\bpi_{s_{i,1}} \cdot f\qty(\vb*{y}^{(i)}_1 \mid \vb*{b}_1^{(i)}=s_{i,1},\; \text{rest})}{\displaystyle\sum_{m=1}^5 \bpi_{m} \cdot f\qty(\vb*{y}^{(i)}_1 \mid \vb*{b}_1^{(i)}=m, \; \text{rest})} \times
        \mathlarger{\mathlarger{\prod}}_{t=2}^{n_i} \frac{\vb{P}_{s_{i,t-1},s_{i,t}} \cdot f\qty(\vb*{y}^{(i)}_t \mid \vb*{y}^{(i)}_{t-1}, \{\vb*{b}_j^{(i)} = s_{i,j}\}_{j=1}^t, \; \text{rest})}{\displaystyle\sum_{m=1}^5\vb{P}_{s_{i,t-1},m} \cdot f\qty(\vb*{y}^{(i)}_t \mid \vb*{y}^{(i)}_{t-1}, \{\vb*{b}_j^{(i)} = s_{i,j}\}_{j=1}^{t-1}, \vb*{b}_t^{(i)} = m, \; \text{rest})}
\end{align*}
}%
Additionally, the acceptance ratio for the MH sampling in Algorithm \ref{chap3:alg:statesamp2} does not require computing the full joint posterior distribution. It can be shown (for $2 < p < n_i$) that the MH acceptance ratio between the proposed state sequence, $\{\bb_j^{(i)} = s^*_{i,j}\}_{j=1}^{n_i}$, and the current state sequence, $\{\bb_j^{(i)} = s_{i,j}\}_{j=1}^{n_i}$, simplifies to
\begin{align*}
    MH_{\text{ratio}} &= \qty{\prod_{t = k_{\text{max}}-1}^{k_{\text{max}}+1} \frac{\vb{P}_{s^*_{i,t-1},s^*_{i,t}}}{\vb{P}_{s_{i,t-1},s_{i,t}}}}\cdot \qty{\prod_{t = k_{\text{max}}-1}^{n_i} \frac{f\qty(\vb*{y}^{(i)}_t \mid \vb*{y}^{(i)}_{t-1}, \{\vb*{b}_j^{(i)} = s^*_{i,j}\}_{j=1}^t, \; \text{rest})}{f\qty(\vb*{y}^{(i)}_t \mid \vb*{y}^{(i)}_{t-1}, \{\vb*{b}_j^{(i)} = s_{i,j}\}_{j=1}^t, \; \text{rest})}}\\
    &\times \qty{\prod_{t = k}^{k_{\text{max}}-2} \frac{\sum_{m=1}^5\vb{P}_{s^*_{i,t-1},m} \cdot f\qty(\vb*{y}^{(i)}_t \mid \vb*{y}^{(i)}_{t-1}, \{\vb*{b}_j^{(i)} = s^*_{i,j}\}_{j=1}^{t-1}, \vb*{b}_t^{(i)} = m, \; \text{rest})}{\sum_{m=1}^5\vb{P}_{s_{i,t-1},m} \cdot f\qty(\vb*{y}^{(i)}_t \mid \vb*{y}^{(i)}_{t-1}, \{\vb*{b}_j^{(i)} = s_{i,j}\}_{j=1}^{t-1}, \vb*{b}_t^{(i)} = m, \; \text{rest})}} \\
    &\times \frac{|\mathcal{B}^{(2)}_{s_{i,k_{\text{max}}-2}^*, s_{i,k_{\text{max}}+1}^*}|}{|\mathcal{B}^{(2)}_{s_{i,k_{\text{max}}-2}, s_{i,k_{\text{max}}+1}}|},
\end{align*}
where each component above has already been computed from the derivation of the proposal distribution.

\newpage

\section{Alternative State-Sampling Routines} \label{chap3:app:altSamp}

\noindent \fbox{\textit{Approach (A)}}
{
\spacingset{1}
\small
\[
\begin{split}
    &q\qty(\{\vb*{b}_j^{(i)} = s_{i,j}\}_{j=k}^{k+p-1} \mid \{\vb*{b}_j^{(i)} = s_{i,j}\}_{j\notin \{k,\dots, k+p-1\}},\; \text{rest})\\
    &\hspace{1.5in}= \begin{cases}
        |\mathcal{B}^{(p)}_{\boldsymbol{\cdot},\; s_{i,k+p}}|^{-1} &,\; k=1\\
        |\mathcal{B}^{(p)}_{s_{i,k-1},\; s_{i,k+p}}|^{-1} &,\; k\in\{2,\hdots, n_i - p\}\\
        |\mathcal{B}^{(p)}_{s_{i,k-1},\; \boldsymbol{\cdot}}|^{-1} &,\; k=n_i - p+1
    \end{cases}
\end{split}
\]
}%

\noindent \fbox{\textit{Approach (B)}}
{
\spacingset{1}
\small
\[
\begin{split}
    &q\qty(\{\vb*{b}_j^{(i)} = s_{i,j}\}_{j=k}^{k+p-1} \mid \{\vb*{b}_j^{(i)} = s_{i,j}\}_{j\notin \{k,\dots, k+p-1\}},\; \text{rest})\\
    &\hspace{0.1in}= \begin{cases}
        \frac{\bpi_{s_{i,1}} \cdot f(\vb*{y}^{(i)}_1 \mid \vb*{b}_1^{(i)}=s_{i,1},\; \text{rest}) {\displaystyle\prod_{t = 2}^{n_i}} \vb{P}_{s_{i,t-1},s_{i,t}} \cdot f\qty(\vb*{y}^{(i)}_t \mid \vb*{y}^{(i)}_{t-1}, \{\vb*{b}_j^{(i)} = s_{i,j}\}_{j=1}^t, \; \text{rest})}{{\displaystyle\sum_{\Tilde{\vb*{s}} \in \mathcal{B}^{(p)}_{\boldsymbol{\cdot}, s_{i,k+p}}}} \bpi_{\Tilde{s}_{1}} \cdot f(\vb*{y}^{(i)}_1 \mid \vb*{b}_1^{(i)}=\Tilde{s}_{1},\; \text{rest}) {\displaystyle\prod_{t = 2}^{n_i}} \vb{P}_{\Tilde{s}_{t-1},\Tilde{s}_{t}} \cdot f\qty(\vb*{y}^{(i)}_t \mid \vb*{y}^{(i)}_{t-1}, \{\vb*{b}_j^{(i)} = \Tilde{s}_{j}\}_{j=1}^t, \; \text{rest})} &,\; k = 1 \\
        \frac{{\displaystyle\prod_{t = k}^{n_i}} \vb{P}_{s_{i,t-1},s_{i,t}} \cdot f\qty(\vb*{y}^{(i)}_t \mid \vb*{y}^{(i)}_{t-1}, \{\vb*{b}_j^{(i)} = s_{i,j}\}_{j=1}^t, \; \text{rest})}{{\displaystyle\sum_{\Tilde{\vb*{s}} \in \mathcal{B}^{(p)}_{s_{i,k-1},\; s_{i,k+p}}}} {\displaystyle\prod_{t = k}^{n_i}} \vb{P}_{\Tilde{s}_{t-1},\Tilde{s}_{t}} \cdot f\qty(\vb*{y}^{(i)}_t \mid \vb*{y}^{(i)}_{t-1}, \{\vb*{b}_j^{(i)} = \Tilde{s}_{j}\}_{j=1}^t, \; \text{rest})} &,\; k\in\{2,\hdots, n_i - p\}\\
        \frac{{\displaystyle\prod_{t = k}^{n_i}} \vb{P}_{s_{i,t-1},s_{i,t}} \cdot f\qty(\vb*{y}^{(i)}_t \mid \vb*{y}^{(i)}_{t-1}, \{\vb*{b}_j^{(i)} = s_{i,j}\}_{j=1}^t, \; \text{rest})}{{\displaystyle\sum_{\Tilde{\vb*{s}} \in \mathcal{B}^{(p)}_{s_{i,k-1},\; \boldsymbol{\cdot}}}} {\displaystyle\prod_{t = k}^{n_i}} \vb{P}_{\Tilde{s}_{t-1},\Tilde{s}_{t}} \cdot f\qty(\vb*{y}^{(i)}_t \mid \vb*{y}^{(i)}_{t-1}, \{\vb*{b}_j^{(i)} = \Tilde{s}_{j}\}_{j=1}^t, \; \text{rest})} &,\; k=n_i-p+1
    \end{cases}
\end{split}
\]
}%

\newpage

\section{Parameterization for Simulation Study}\label{chap3:app:simulationPar}
{\small
\begin{align*}
    &\bbeta = \mqty(0.4322\\ -0.7361\\  1.8589\\  0.0361),  
    \quad \Tilde{\balpha}_* = \mqty(-0.8429  & 6.6528 & -9.5501&  0.7762\\ 0.7443 & -6.9619 & 8.0447 & -0.7595\\ 0.0919 & 0.0330  & 2.2708 &  0.2091\\ -0.1177 & -1.3514 & -1.4781 & -0.0061),
\end{align*}
\begin{align*}
    &\bA_1 = \text{diag}(0.1652, 0.9707, 0.8804, 0.8664), 
    \quad \bA_2 = \text{diag}(0.1412, 0.0041, 0.0020, 0.8122),\\
    &\bA_3 = \text{diag}(0.5263, 0.2666, 0.0068, 0.9035), 
    \quad \bA_4 = \text{diag}(0.2314, 0.0009, 0.0013, 0.7314),\\
    &\bA_5 = \text{diag}(0.1029, 0.3675, 0.1392, 0.2065),
\end{align*}
\begin{align*}
    &\bR = \mqty(0.49852 & 0.00097 & 0.00236 & 0.00001\\
                 0.00097 & 5.61452 & 1.51487 & 0.00083\\
                 0.00236 & 1.51487 & 10.55849 & -0.00111\\
                0.00001 & 0.00083 & -0.00111 &  0.49784),
\end{align*}
\begin{align*}
    &\bzeta = \mqty(-2.9732,-2.1213,1.9556,0.6334,1.0973,0.1456,0.3555,-1.0543,0.7188,-0.1747,-1.9821,-0.4705\\
    0.5475,0.1003,-0.5696,0.0869,-1.3116,-0.1631,-0.8327,0.5962,0.2341,-0.6008,0.1264,-0.1105)
\end{align*}
\begin{align*}
    &\text{diag}(\bUpsilon_\alpha) = (0.1437,0.1289,0.9086,0.9075,33.012,35.8889,26.9045,20.0647, \\
    &\hspace{0.85in}\;54.0599,42.9315,43.4438,26.2896,0.1466,0.1356,1.0752,1.1122),
\end{align*}
\begin{align*}
&\bomega = (-1.458,1.307,1.453,-1.046,-1.721,0.95,0.982,-1.914,0.435,0.742,-0.835,-1.516,\\
                &\quad 1.666,-1.709,1.482,1.511,-1.133,-1.251,-0.76,-1.518,1.497,-0.596,1.915,-0.311,\\
                &\quad 0.274,-1.436,-0.606,-0.688,-0.251,-0.941,1.589,0.816,-1.299,-1.368,-1.27,-1.292,\\
                &\quad -1.486,1.051,1.474,1.063,-1.095,1.22,-1.33,-1.562,-1.136,1.88,-1.467,0.703,\\
                &\quad -1.332,0.928,-1.534,-0.733,-1.26,1.616,1.272,-1.489,-1.158,-1.494,-1.384,-0.874,\\
                &\quad -1.494,-0.341,-1.565,-1.229,-1.343,0.574,0.82,0.924,-1.187,-1.351,-0.715,0.018,\\
                &\quad -0.447,1.918,-0.997,-1.204,-1.532,-1.09,1.44,-1.604,-1.515,-1.589,-1.32,-1.341)
\end{align*}
}%

\newpage

\section{Simple Simulation Study}\label{chap3:app:simpleSim}
The data for this simple simulation study are generated according to the description in Section \ref{chap3:simDataGen} of the manuscript; however, the data generating mechanism here is simpler.

First, the state-space for this example contains only \textit{three} states and the state sequences are assumed to follow a Markov process with transition probability matrix defined as 
$$\vb{P} = \mqty(0.8808 & 0.1192 & 0\\ 0 & 0.8808 & 0.1192\\ 0.1543 & 0.1543 & 0.6914).$$
Mimicking Section \ref{chap3:simDataGen} of the manuscript, for each subject $i \in \{1,2,\hdots, 500\}$, we generate a state sequence of length $m_i + n_i$ (namely, $\vb*{b}_{long}^{(i)}$) that always starts in state 1, but only the last $n_i$ time points are used as the true state sequence (namely, $\vb*{b}^{(i)}$). Then, for every $i$, let $m_i \sim \text{uniform}\{0,1,\hdots, 50\}$ and $n_i \sim \text{Poisson}(\lambda = 100)$.

Next, we simplify the true data generating conditional response model from the simulation in Section \ref{chap3:sec:sim} of the manuscript. Notably, the simulation here no longer contains random effects nor covariates, and the mean process is no longer autoregressive. The only similarity to the simulation study in the manuscript is the state-dependence structure. Thus, the data generating mechanism for the conditional response is defined as:
{\small
\begin{align*}
    &\by^{(i)}_1 \mid \bb_1^{(i)} = s_{i,1}, \balpha, \bR  &&\sim \text{N}_4\qty(g(\balpha, \vb*{b}^{(i)}_1),\; \bR)\\
    &\by^{(i)}_k \mid \{\bb_j^{(i)} = s_{i,j}\}_{j=1}^k, \balpha, \bR  &&\sim \text{N}_4\qty(g(\balpha, \vb*{b}^{(i)}_1) + \Big[\sum_{j=2}^{k}\mathbf{1}\{\bb^{(i)}_{j}=2\}\Big] \balpha_{\cdot, 2} + \Big[\sum_{j=2}^{k}\mathbf{1}\{\bb^{(i)}_{j}=3\}\Big] \balpha_{\cdot, 3},\; \bR),
\end{align*}
}%
for $k \in \{2,3,\hdots, n_i\}$ and $s_{i,j}\in\{1,2,3\}$ for all $i$ and for all $j \in \{1,\hdots, n_i\}$. Define $g(\balpha, \bb^{(i)}_1) := \balpha_{\cdot,1} + t_2^{(i)} \balpha_{\cdot,2} + t_3^{(i)} \balpha_{\cdot,3}$ and $t_l^{(i)} := \sum_{j = 1}^{m_i+1} \mathbf{1}\{\vb*{b}_{long,\; j}^{(i)} = l\}$ for $l \in \{2,3\}$. The true parameter values are then given by
\begin{align*}
    \vb*{\alpha} &= \mqty(50 & -5 & 5 \\ 100 & 10 & -10 \\ 100 & -10 & 10\\ 50 & 5 & -5), & \vb*{R} = \text{diag}(4,4,4,4).
\end{align*}

The purpose of this simulation study is to illustrate the inferential consequences of \textit{not} accounting for pre-ICU-admission physiological changes, and verifying that our approach (as described in Section \ref{chap3:subsec:initAdj} of the manuscript) works for handling these pre-ICU-admission physiological changes. We fit two models (\textit{Model A} and \textit{Model B}). For both, we assume the latent state process is defined by a Markov process, and we will learn the discrete, initial state probability distribution $\vb*{\pi}$ and the $3\times 3$ transition probability matrix $\vb{P}$. These two models vary are in their definitions of the conditional response:
{\small
\begin{align*}
    &\fbox{\textbf{Model A}} &&\\
    &\by^{(i)}_k \mid \{\bb_j^{(i)} = s_{i,j}\}_{j=1}^k, \balpha, \bR  &&\sim \text{N}_4\qty(\balpha_{\cdot, 1} + \Big[\sum_{j=1}^{k}\mathbf{1}\{\bb^{(i)}_{j}=2\}\Big] \balpha_{\cdot, 2} + \Big[\sum_{j=1}^{k}\mathbf{1}\{\bb^{(i)}_{j}=3\}\Big] \balpha_{\cdot, 3},\; \bR),\\
    & \text{for } k\in\{1,2,\hdots, n_i\}.&&\\
    &\fbox{\textbf{Model B}} &&\\
    &\by^{(i)}_1 \mid \bgamma^{(i)}, \bR &&\sim \text{N}_4(\bgamma^{(i)},\; \vb*{R})\\
    &\by^{(i)}_k \mid \{\bb_j^{(i)} = s_{i,j}\}_{j=2}^k, \balpha_*, \bgamma^{(i)}, \bR &&\sim \text{N}_4\Bigg(\bgamma^{(i)} + \Big[\sum_{j=2}^{k}\mathbf{1}\{\bb^{(i)}_{j}=2\}\Big] \balpha^*_{\cdot, 1} + \Big[\sum_{j=2}^{k}\mathbf{1}\{\bb^{(i)}_{j}=3\}\Big] \balpha^*_{\cdot, 2},\; \bR\Bigg),
\end{align*}
}%
for $k\in\{2,\hdots, n_i\}$ where $\bgamma^{(i)} \mid \by^{(i)}_{1}, \vb*{G} \sim \text{N}_4(\by^{(i)}_1,\; \vb*{G})$, $\vb*{G}$ is a $4\times 4$ covariance matrix, and $\vb*{\alpha}^*$ is a $4\times 2$ matrix of slope coefficients for the expected change in the four-dimensional response as a result of being in states 2 or 3, respectively for each column.

Model A serves as a naive strategy that does \textit{not} account for pre-ICU-admission physiological changes because it assumes that if the initial state is not state 1, then the subject has \textit{only} been in either state 2 or state 3 for one time instance. Model B, however, follows from the strategy discussed in Section \ref{chap3:subsec:initAdj} of the manuscript. Notice that from the data generating mechanism, we have no way of learning the true intercept term $\balpha_{\cdot, 1}$ without knowing $t_2^{(i)}$ and $t_3^{(i)}$, for all $i$. In lieu of this, Model B has no intercept term to learn; instead we learn $\balpha_{\cdot, 2}$ and $\balpha_{\cdot, 3}$ by estimating $\balpha^*_{\cdot, 1}$ and $\balpha^*_{\cdot, 2}$, respectively.

\begin{figure}[!htb]
    \centering
    \spacingset{1}
    \includegraphics[page = 1,trim={2.5in 0 2.5in 0},clip, height=4.5in]{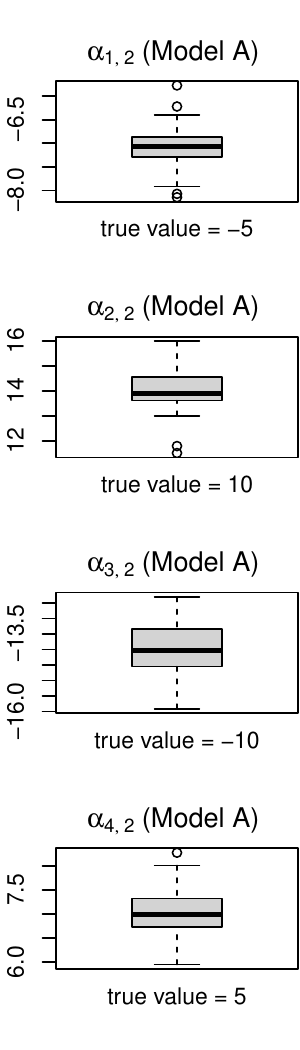}
    \includegraphics[page = 1,trim={0 0 4.8in 0},clip, height=4.5in]{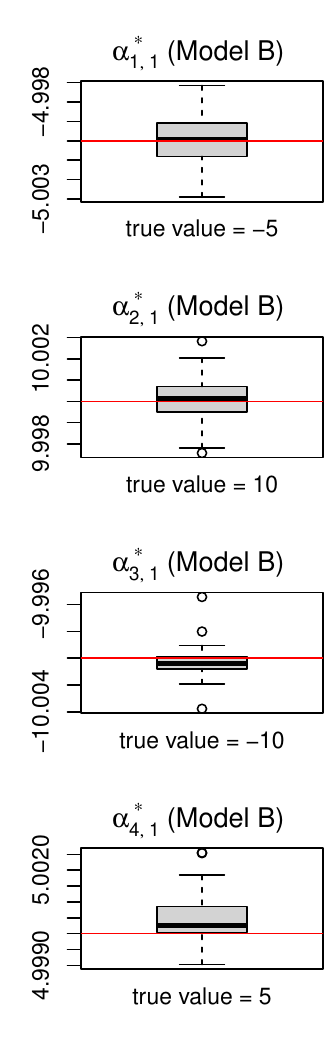}
    \includegraphics[page = 1,trim={4.5in 0 0 0},clip, height=4.5in]{Plots/box_plot_1.pdf}
    \includegraphics[page = 1,trim={2.5in 0 2.5in 0},clip, height=4.5in]{Plots/box_plot_0.pdf}
    \caption{\footnotesize Box plots of the posterior means across 25 simulated datasets. These four columns specifically compare the estimated slope coefficients corresponding to state 2 (columns 1 and 2) and state 3 (columns 3 and 4) using Model A and Model B to fit the data. The \textit{true} parameter value is provided on the x-axis, as well as marked by a red horizontal line, in each plot.}
    \label{chap3:app:figboxplot}
\end{figure}

We fit Models A and B to 25 distinct datasets by running a Metropolis-within-Gibbs MCMC sampling routine for 10,000 steps, discarding the first 5,000 steps for burnin. Figure \ref{chap3:app:figboxplot} presents box plots of the posterior means (from fitting Models A and B) for the slope coefficients corresponding to the effect of state 2 and state 3 on the mean of the response. We see from Figure \ref{chap3:app:figboxplot} that our proposed approach (Model B) correctly centers the posterior means around the true parameter value, while Model A exhibits biased and more variable estimation of these slopes. We can attribute these estimation results from Model A to the fact that it does not properly account for the state changes that occur before the first observation. 

\begin{table}[!htb]
    \centering\footnotesize
    \begin{tabular}{|l| c c c c|}
    \hline
    & Q1 & Median & Mean & Q3\\
    \hline
    Model A & 0.6427 & 0.7092 & 0.7082 & 0.7753\\
    Model B & 1.0000 & 1.0000 & 0.9957 & 1.0000 \\
    \hline
    \end{tabular}
    \caption{\footnotesize Summary of how Model A and Model B perform with respect to identifying the latent state sequences across all subjects, for one of the 25 simulated datasets.}
    \label{chap3:app:tabAcc}
\end{table}

Next, we can compare how Model A and Model B perform with respect to accurately identifying the true underlying state sequence for each subject (``accuracy'' defined as the proportion of time instances where the posterior modal state correctly corresponds to the true state). Table \ref{chap3:app:tabAcc} provides a summary of each models' performance. We again see that our proposed approach (Model B) outperforms Model A, this time with respect to learning the underlying latent state sequences for each subject in the data.

Trace plots and the remaining box plots for the other model parameters can be found by either emailing the author or visiting \url{https://ebkendall.github.io/research.html}.

\newpage 

\section{Estimated Medication Effects}\label{chap3:app:medEffect}
\vspace{-0.5cm}

\begin{table}[H]
    \scriptsize
    \centering
    \begin{tabular}{|c|l l l|l l l|}
    \hline
    & \multicolumn{3}{|c|}{Heart Rate Effect} & \multicolumn{3}{|c|}{MAP Effect}\\
    Medication & Exp.  & Cont. & Disc. & Exp. & Cont. & Disc.\\
    \hline
 ANGIOTENSIN  &     &  --  &  --  &  $\uparrow$   &  $1.627^*$  &  --  \\ 
 EPHEDRINE  &   $\uparrow$  &  --  &  $1.912^*$  &   $\uparrow$  &  --  &  $0.801^*$  \\ 
 EPINEPHRINE  &  $\uparrow$  &  $0.517^*$  &  $1.505^*$  & $\uparrow$  &  $0.754^*$  &  $1.92^*$  \\ 
 NOREPINEPHRINE  &  $\uparrow$   &  $0.971^*$  &  --  &  $\uparrow$   &  $1.234^*$  &  --  \\ 
 PHENYLEPHRINE  &  $\downarrow$ &  $-1.1^*$  &  $-0.302^*$  &  $\uparrow$ &  $1.003^*$  &  $0.891^*$  \\ 
 VASOPRESSIN  &     &  --  &  --  &  $\uparrow$   &  $1.968^*$  &  --  \\ 
 DOBUTAMINE  &  $\uparrow$   &  $1.691^*$  &  --  &  $\downarrow$   &  $-0.817^*$  &  --  \\ 
 DOPAMINE  &  $\uparrow$   &  $1.504^*$  &  --  &  $\uparrow$   &  $1.276^*$  &  --  \\ 
 MILRINONE  &  $\uparrow$   &  $1.182^*$  &  --  &  $\downarrow$   &  $-1.281^*$  &  --  \\ 
 ATROPINE  &   $\uparrow$  &  --  &  $0.879^*$  &   $\uparrow$  &  --  &  $1.436^*$  \\ 
 CALCIUM CHLORIDE  &     &  --  &  --  &  $\uparrow$ &  $0.914^*$  &  $0.014$  \\ 
 CALCIUM GLUCONATE  &     &  --  &  --  &   $\uparrow$  &  --  &  $0.56^*$  \\ 
 ISOPROTERENOL  &  $\uparrow$   &  $1.507^*$  &  --  &  $\downarrow$   &  $-1.496^*$  &  --  \\ 
 SODIUM BICARBONATE  &     &  --  &  --  &  $\uparrow$   &  $1.15^*$  &  --  \\ 
 ALBUMIN  &  $\downarrow$   &  $-1.612^*$  &  --  &  $\uparrow$   &  $1.522^*$  &  --  \\ 
 DILTIAZEM  &  $\downarrow$ &  $-1.464^*$  &  $-0.69^*$  &  $\downarrow$ &  $-1.227^*$  &  $-1.313^*$  \\ 
 ADENOSINE  &   $\downarrow$  &  --  &  $-0.946^*$  &     &  --  &  --  \\ 
 AMIODARONE  &  $\downarrow$ &  $-1.925^*$  &  $-1.119^*$  & $\downarrow$  &  $-1.403^*$  &  $-1.461^*$  \\ 
 AMLODIPINE  &     &  --  &  --  &   $\downarrow$  &  --  &  $-0.821^*$  \\ 
 ATENOLOL  &   $\downarrow$  &  --  &  $-1.502^*$  &   $\downarrow$  &  --  &  $-1.503^*$  \\ 
 BISOPROLOL  &   $\downarrow$  &  --  &  $-1.186^*$  &   $\downarrow$  &  --  &  $-1.332^*$  \\ 
 CAPTOPRIL  &     &  --  &  --  &   $\downarrow$  &  --  &  $-1.509^*$  \\ 
 CARVEDILOL  &   $\downarrow$  &  --  &  $-0.752^*$  &   $\downarrow$  &  --  &  $-0.794^*$  \\ 
 CLEVIDIPINE  &     &  --  &  --  &  $\downarrow$   &  $-1.098^*$  &  --  \\ 
 CLONIDINE  &     &  --  &  --  &   $\downarrow$  &  --  &  $-1.06^*$  \\ 
 DIGOXIN  &   $\downarrow$  &  --  &  $-0.626^*$  &     &  --  &  --  \\ 
 ESMOLOL  &  $\downarrow$ &  $-1.722^*$  &  $-1.268^*$  &  $\downarrow$ &  $-1.354^*$  &  $-1.571^*$  \\ 
 ISOSORBIDE DINITRATE  &     &  --  &  --  &   $\downarrow$  &  --  &  $-1.621^*$  \\ 
 ISOSORBIDE MONONITRATE  &     &  --  &  --  &   $\downarrow$  &  --  &  $-1.493^*$  \\ 
 LABETALOL  &  $\downarrow$ &  $-1.531^*$  &  $-0.78^*$  &  $\downarrow$  &  $-1.528^*$  &  $-1.405^*$  \\ 
 LISINOPRIL  &     &  --  &  --  &   $\downarrow$  &  --  &  $-0.387^*$  \\ 
 LOSARTAN  &     &  --  &  --  &   $\downarrow$  &  --  &  $-1.401^*$  \\ 
 METOPROLOL  &   $\downarrow$  &  --  &  $-0.354^*$  &   $\downarrow$  &  --  &  $-0.455^*$  \\ 
 METOPROLOL SUCCINATE  &   $\downarrow$  &  --  &  $-1.489^*$  &   $\downarrow$  &  --  &  $-1.238^*$  \\ 
 METOPROLOL TARTRATE  &   $\downarrow$  &  --  &  $-1.102^*$  &   $\downarrow$  &  --  &  $-1.045^*$  \\ 
 NICARDIPINE  &     &  --  &  --  &  $\downarrow$   &  $-1.43^*$  &  --  \\ 
 NIFEDIPINE  &     &  --  &  --  &   $\downarrow$  &  --  &  $-1.18^*$  \\ 
 NIMODIPINE  &     &  --  &  --  &   $\downarrow$  &  --  &  $-1.497^*$  \\ 
 NITROGLYCERIN  &  $\uparrow$  &  $0.959^*$  &  $1.502^*$  &  $\downarrow$ &  $-1.339^*$  &  $-1.584^*$  \\ 
 NITROPRUSSIDE  &  $\uparrow$   &  $1.397^*$  &  --  &  $\downarrow$   &  $-1.564^*$  &  --  \\ 
 SILDENAFIL  &     &  --  &  --  &   $\downarrow$  &  --  &  $-1.319^*$  \\ 
 DEXMEDETOMIDINE  &  $\downarrow$   &  $-0.9^*$  &  --  &  $\downarrow$   &  $-1.328^*$  &  --  \\ 
 KETAMINE  &  $\uparrow$ &  $0.751^*$  &  $0.321^*$  &     &  --  &  --  \\ 
 PROPOFOL  &     &  --  &  --  & $\downarrow$  &  $-1.082^*$  &  $-1.043^*$  \\
    \hline
    \end{tabular}
    \caption{\footnotesize Posterior median medication effects from the real data analysis. The $^*$ indicates that the empirical 95\% credible intervals exclude zero. The expected effect of the medication on heart rate and MAP is provided as a directional arrow where $\uparrow$ indicates an upward or positive effect, while $\downarrow$ indicates a downward or negative effect.}
    \label{chap3:tab:medEffect}
\end{table}

\section{Trace Plots and Box Plots (Simulation)} \label{chap3:app:simulationTrace}
For clarity and conciseness, \texttt{pdf} documents of the trace plots for the 100 simulations and the box plots of the 100 posterior medians can be found by either emailing the author or visiting \url{https://ebkendall.github.io/research.html}.

\section{Trace Plots (Real Data Analysis)}\label{chap3:app:traceRealData}
For clarity and conciseness, \texttt{pdf} documents of the trace plots for the real data analysis can be found by either emailing the author or visiting \url{https://ebkendall.github.io/research.html}.

\section{Chart Plots (Test Set)} \label{chap3:app:chartTest}
\subsection{Subject A}
\begin{figure}[!htb]
    \centering
    \spacingset{1}
    \includegraphics[page = 1, width=\linewidth]{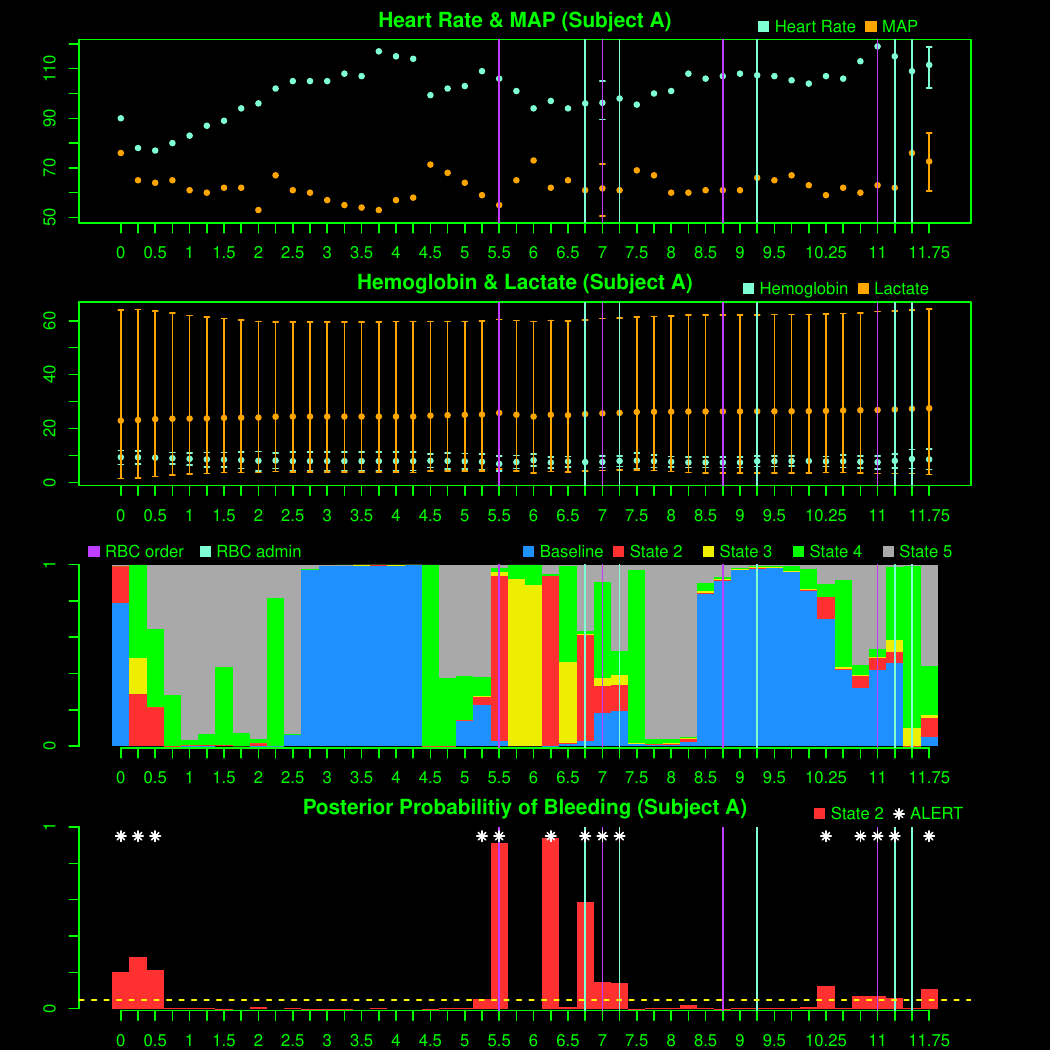}
    \caption{\footnotesize The top two panels correspond to the longitudinal vital measurements. The points with the error bars correspond to missing values; the error bars are empirical 95\% credible intervals for the imputed response values. The third panel depicts the discrete posterior probability distributions of the latent states, at each time point. The bottom panel is the posterior probability of state 2 at each time point, and the yellow dashed line represents the threshold $\hat{c} = 0.0465$ determined from Section \ref{chap3:subsec:postInt}. The white stars indicate that the posterior probability of state 2 exceeds the threshold. The purple and turquoise vertical lines represent RBC transfusion order and administration times, respectively.}
    \label{chap3:fig:chartPlot1}
\end{figure}

Figure \ref{chap3:fig:chartPlot1} demonstrates the course for a 63-year-old male who underwent liver transplant for alcoholic cirrhosis. He had been admitted for more than 30 days prior to transplant. Intraoperatively, the case proceeded well. He required 8 units of cryoprecipitate, 16 units of fresh frozen plasma (FFP), 4 units of platelets, and 11 units of RBCs during the procedure. Postoperatively, he continued to require transfusion of blood products as well as vasopressor support with two agents. Due to this, he was taken back to the operating room within 12 hours where a large amount of clot was evacuated and areas of bleeding were controlled. 

The model predicted a bleeding event as can be seen in Figure \ref{chap3:fig:chartPlot1} which was confirmed in this patient and ultimately required return to the operating room for resolution.  

\subsection{Subject B}
\begin{figure}[!htb]
    \centering
    \spacingset{1}
    \includegraphics[page = 2, width=\linewidth]{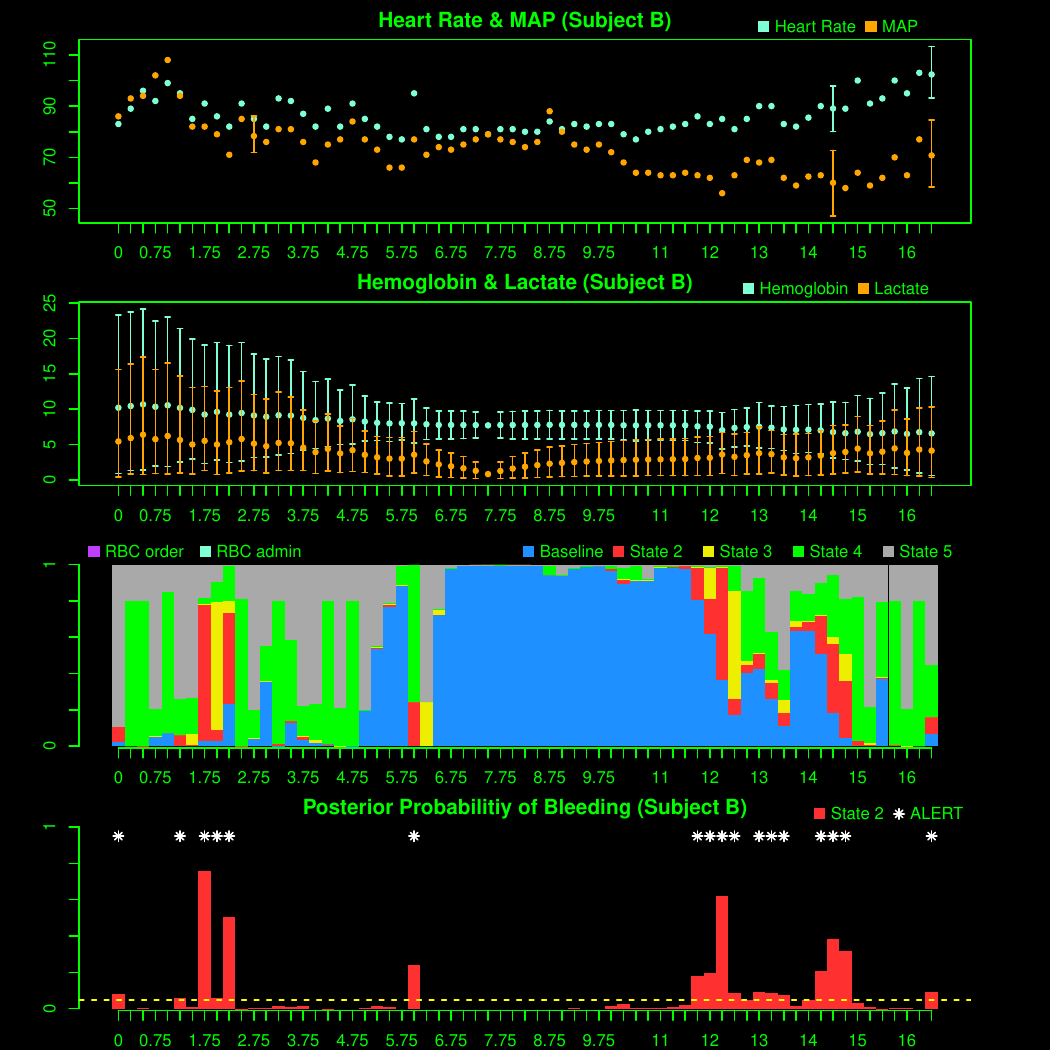}
    \caption{\footnotesize The top two panels correspond to the longitudinal vital measurements. The points with the error bars correspond to missing values; the error bars are empirical 95\% credible intervals for the imputed response values. The third panel depicts the discrete posterior probability distributions of the latent states, at each time point. The bottom panel is the posterior probability of state 2 at each time point, and the yellow dashed line represents the threshold $\hat{c} = 0.0465$ determined from Section \ref{chap3:subsec:postInt}. The white stars indicate that the posterior probability of state 2 exceeds the threshold. The purple and turquoise vertical lines represent RBC transfusion order and administration times, respectively.}
    \label{chap3:fig:chartPlot2}
\end{figure}

Figure \ref{chap3:fig:chartPlot2} demonstrates the course for a medically complex 65-year-old woman transferred from her local hospital for management of an enterocutaneous fistula. During her stay at the local hospital, she underwent surgery to manage a small bowel obstruction which required resection of portions of her small bowel and colon and the creation of a colostomy. Unfortunately, her course there was complicated by sepsis and multiorgan failure including respiratory failure requiring mechanical ventilation, renal failure requiring continuous dialysis, liver failure, and delirium. During the ICU course outlined in Figure \ref{chap3:fig:chartPlot2} following transfer to Mayo Clinic, she required vasopressor support for hypotension but did not develop a lactic acidosis. No bleeding was identified. She was transfused during the timeline in Figure \ref{chap3:fig:chartPlot2}, but for a subacute anemia related to phlebotomy, critical illness, liver and renal failure rather than hemorrhage. 

Although Figure \ref{chap3:fig:chartPlot2} suggested a likely bleeding event occurring during the time depicted, the patient did not suffer a bleeding event. This may represent over-detection of hemorrhage by the model related to heart rate and MAP trends.

\begin{figure}[!htb]
    \centering
    \spacingset{1}
    \includegraphics[page = 4, width=\linewidth]{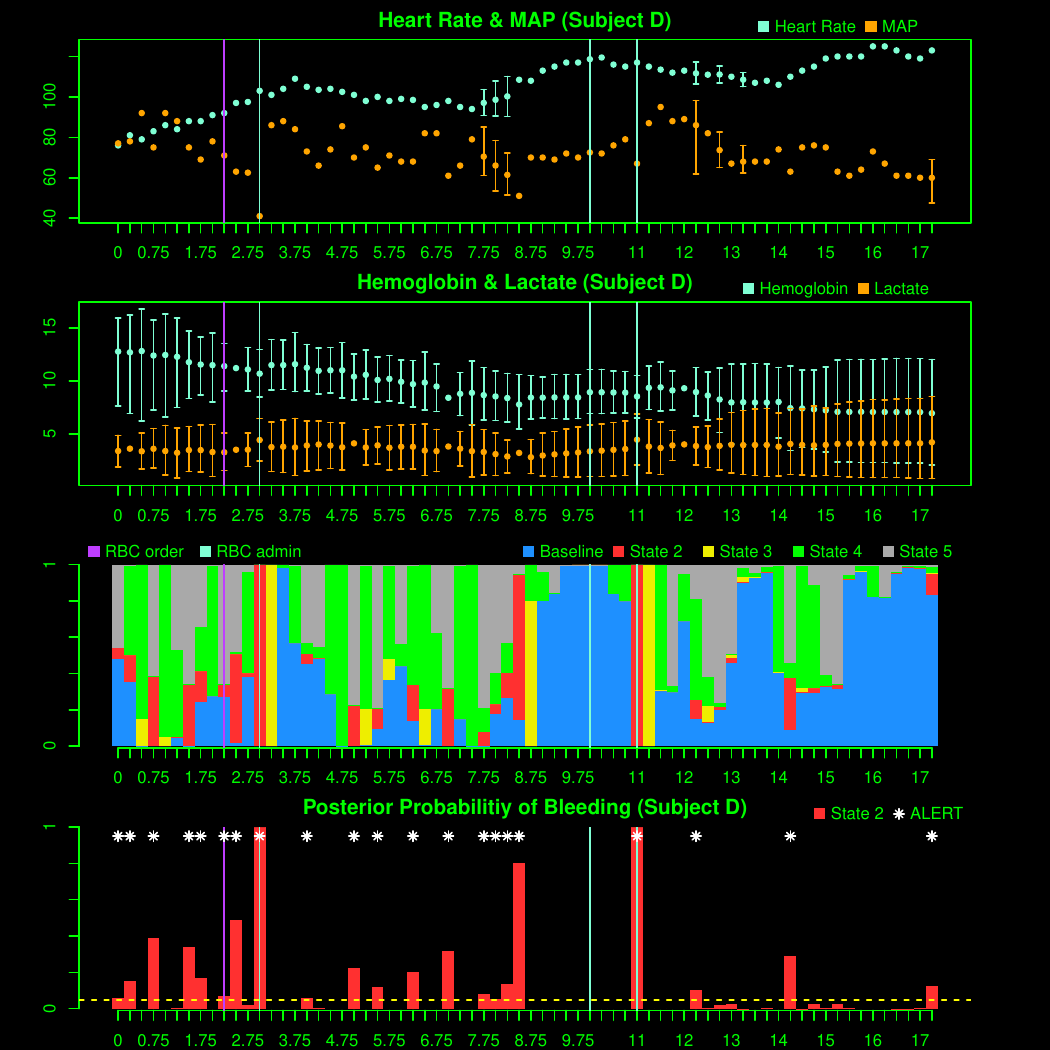}
    \caption{\footnotesize The top two panels correspond to the longitudinal vital measurements. The points with the error bars correspond to missing values; the error bars are empirical 95\% credible intervals for the imputed response values. The third panel depicts the discrete posterior probability distributions of the latent states, at each time point. The bottom panel is the posterior probability of state 2 at each time point, and the yellow dashed line represents the threshold $\hat{c} = 0.0465$ determined from Section \ref{chap3:subsec:postInt}. The white stars indicate that the posterior probability of state 2 exceeds the threshold. The purple and turquoise vertical lines represent RBC transfusion order and administration times, respectively.}
    \label{chap3:fig:chartPlot4}
\end{figure}

\subsection{Subject D}
Figure \ref{chap3:fig:chartPlot4} demonstrates the course for a medically complex 76-year-old female who presented with right lower extremity non-healing wounds. She had previously undergone bilateral common femoral endarterectomies and retrograde iliac stents earlier in the year for critical limb ischemia.  She developed thrombosis of the right common femoral artery and critical stenosis of the left common femoral artery.  The wound on her right leg had enlarged in size and she was having severe rest pain. She went to surgery for revascularization via stenting or bypass. An endovascular approach was performed, however, before the intervention could be done, the patient became unstable from an intraoperative bleeding event. This was treated with a stent at the aortic bifurcation and the patient required a laparotomy due to elevated airway pressures, decreased urine output, and abdominal distention. The hematoma was evacuated and her abdomen was left open. During the case she received 14 units of packed red blood cells (PRBC), 3 units of platelets, 5 units of FFP, and 4 units of cryoprecipitate. On arrival to the ICU she was hemodynamically stable, but over the course of the day, she required additional blood products, had bright red drain output, required vasopressors, and had an increase in lactate. She returned to the operating room due to concern for limb ischemia and was found to have hemoperitoneum and ischemic bowel as well. She underwent bowel resection and thrombectomy. Bleeding sites were treated. 

The model was able to detect that the patient was bleeding internally. 

\subsection{Subject E}
\begin{figure}[!htb]
    \centering
    \spacingset{1}
    \includegraphics[page = 5, width=\linewidth]{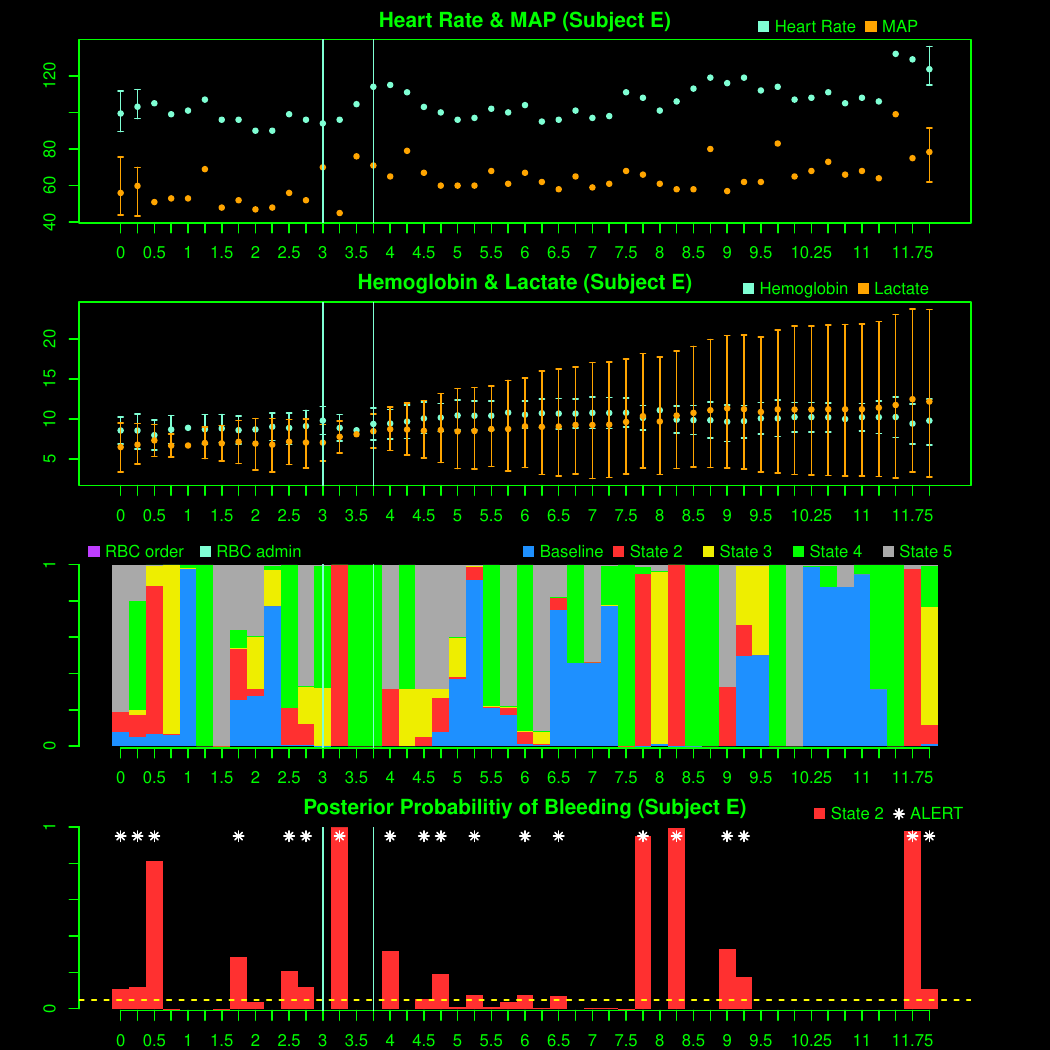}
    \caption{\footnotesize The top two panels correspond to the longitudinal vital measurements. The points with the error bars correspond to missing values; the error bars are empirical 95\% credible intervals for the imputed response values. The third panel depicts the discrete posterior probability distributions of the latent states, at each time point. The bottom panel is the posterior probability of state 2 at each time point, and the yellow dashed line represents the threshold $\hat{c} = 0.0465$ determined from Section \ref{chap3:subsec:postInt}. The white stars indicate that the posterior probability of state 2 exceeds the threshold. The purple and turquoise vertical lines represent RBC transfusion order and administration times, respectively.}
    \label{chap3:fig:chartPlot5}
\end{figure}

Figure \ref{chap3:fig:chartPlot5} relates to a 65-year-old man admitted with septic shock and found to have a small bowel perforation with fecal contamination of his peritoneum. He underwent an exploratory laparotomy, resection of the perforated bowel and temporary abdominal closure. He required significant vasopressor and inotropic support due to septic shock. Approximately one liter of blood loss was recorded for the operation.  Following surgery, he was transferred to the ICU where he received two units of blood to correct anemia from blood loss during the operation. During his ICU stay, there was no concern for occult blood loss. He returned to the operating room multiple times during his hospital admission to re-evaluate the bowel and eventually underwent abdominal closure.

Figure \ref{chap3:fig:chartPlot5} indicates several episodes of state 2 for which the model defines as a bleeding event. Although heart rate was rising, MAP was declining and hemoglobin was trending down, clinically, these time points were periods of post-operative resuscitation of a septic shock patient without occult or other hemorrhage.